\documentclass[sigconf]{acmart}

\usepackage{cite}
\usepackage{amsmath,amsfonts}
\usepackage{algorithmic}
\usepackage{graphicx}
\usepackage{booktabs}
\usepackage{textcomp}
\usepackage{comment}
\usepackage{xcolor}
\usepackage{booktabs}
\usepackage{multirow}
\usepackage{hyperref}
\usepackage{fancyhdr}
\usepackage[T1]{fontenc}
\usepackage[utf8]{inputenc}
\usepackage{listings}
\usepackage{subfig}
\usepackage[export]{adjustbox}
\usepackage{enumitem}
\mathchardef\mhyphen="2D

\newcommand{\ET}[1]{\textcolor{black}{#1}}

\newcommand{\Ra}[1]{\textcolor{black}{#1}}
\newcommand{\Rb}[1]{\textcolor{black}{#1}}
\newcommand{\Rc}[1]{\textcolor{black}{#1}}

\def\BibTeX{{\rm B\kern-.05em{\sc i\kern-.025em b}\kern-.08em
    T\kern-.1667em\lower.7ex\hbox{E}\kern-.125emX}}

\AtBeginDocument{%
  \providecommand\BibTeX{{%
    \normalfont B\kern-0.5em{\scshape i\kern-0.25em b}\kern-0.8em\TeX}}}

\setcopyright{none}
\renewcommand\footnotetextcopyrightpermission[1]{} 
\usepackage{fancyhdr}
\fancypagestyle{mystyle}{%
    \fancyhead[L]{To appear in SIGMOD 2024: Proceedings of the 2024 International Conference on Management of Data}\fancyhead[R]{ } 
}%
\begin{document}

\title{Machine Unlearning in Learned Databases: \\An Experimental Analysis}


\author{Meghdad Kurmanji}
\email{Meghdad.Kurmanji@warwick.ac.uk}
\affiliation{%
  \institution{University of Warwick}
  \country{}
}

\author{Eleni Triantafillou}
\email{etriantafillou@google.com}
\affiliation{%
  \institution{Google DeepMind}
  \country{}
}

\author{Peter Triantafillou}
\email{p.triantafillou@warwick.ac.uk}
\affiliation{%
  \institution{University of Warwick}
  \country{}
}



\begin{abstract}
Machine learning models based on neural networks (NNs) are enjoying ever-increasing attention in the Database (DB) community, both in research and practice. 
However, an important issue has been largely overlooked, namely the challenge of dealing with the inherent, highly dynamic nature of DBs, where data updates are fundamental, highly-frequent operations (unlike, for instance, in ML classification tasks).  
Although some recent research has addressed the issues of maintaining updated NN models in the presence of new data insertions, 
the effects of data deletions (a.k.a., "machine unlearning") remain a blind spot. 
With this work, for the first time to our knowledge, we pose and answer the following key questions:
What is the effect of unlearning algorithms on NN-based DB models? 
How do these effects translate to effects on key downstream DB tasks, such as cardinality/selectivity estimation (SE), approximate query processing (AQP), data generation (DG), and upstream tasks like data classification (DC)?
What metrics should we use to assess the impact and efficacy of unlearning algorithms in learned DBs?
Is the problem of (and solutions for) machine unlearning in DBs different from that of machine learning in DBs in the face of data insertions?
Is the problem of (and solutions for) machine unlearning for DBs different from unlearning in the ML literature?
what are the overhead and efficiency of unlearning algorithms (versus the naive solution of retraining from scratch)?
What is the sensitivity of unlearning on batching delete operations (in order to reduce model updating overheads)?
If we have a suitable unlearning algorithm (forgetting old knowledge), can we combine it with an algorithm handling data insertions (new knowledge) en route to solving the general adaptability/updatability requirement in learned DBs in the face of both data inserts and deletes? 
We answer these questions using a comprehensive set of experiments, various unlearning algorithms, a variety of downstream DB tasks (such as SE, AQP, and DG), and an upstream task (DC),  each with different NNs,  and using a variety of metrics (model-internal, and downstream-task specific) on a variety of real datasets, making this also a first key step towards a benchmark for learned DB unlearning.

\end{abstract}

\keywords{Learned database systems, data deletions, machine unlearning}



\maketitle
\thispagestyle{mystyle}
\pagestyle{plain}

\section{Introduction}
ML has been recently enjoying increasing attention from the DB community. 
ML models are being developed to aid DB systems in performing better on a large variety of tasks, including cardinality/selectivity estimation (SE) \citep{hasan2020deep, yang2019deep, yang2020neurocard, zhu2020flat, wang2020we,factorjoin,negi2021flow}, Approximate Query Processing (AQP) 
\citep{ma2019dbest, ma2021learned, thirumuruganathan2020approximate, hilprecht13deepdb}, 
query cost estimation \citep{siddiqui2020cost, zhi2021efficient}, learned indices \citep{ding2020alex, ding2020tsunami, nathan2020learning, kraska2018case}, Data Generation (DG) \citep{choi2017generating, park2018data, xu2019modeling}, workload forecasting \citep{zhu2019novel}, DB tuning \citep{li2019qtune, van2017automatic}, etc. 

One critical challenge learned DB systems are facing is adapting to the dynamic nature of database systems. 
Analytical DBs face frequent insertions, while transactional DBs face frequent insertions, deletions, and in-place updates (the latter can be modelled as a deletion followed by an insertion). 
The need for updatability/adaptability of the learned DB models is thus a fundamental requirement for learned DBs.
Recently, this has started being addressed in-depth. 
Specifically, \citet{li2022warper} have introduced a method to update learned Cardinality Estimators (CE) when there is a drift in data or workload. Although a step forward, their framework can only update workload-driven models and for CE applications only. 
More recently, \citep{kurmanji2022detect} have provided a solution based on transfer learning and knowledge-distillation to update learned DBs in the presence of data insertions carrying out-of-distribution (OOD) data. These papers clearly show that in the presence of distribution drifts, appropriate methods must be developed for model updating. However, these solutions are limited to data insertions and have not been studied under data deletions. 

In this paper, we initiate the study of data deletion in NN-based learned DBs. Specifically, we study its effect on the learned DB components and inform the community of our findings vis-a-vis lessons learned from related research in ML and from insertion updates in DBs. 
In ML, data deletion materializes as the problem of  "Machine Unlearning". 
This is the problem of removing information related to a part of a dataset from a trained model, without hurting the information about the rest of the data \citep{ginart2019making}. This has also been referred to as ``forgetting'', ``scrubbing'', ``removing'', and ``deletion''. We will use these terms interchangeably. 

Machine unlearning in ML research is motivated by the need to deal with out-of-date, noisy, poisoned, or outlier data. 
Another important reason for machine unlearning is to protect users' privacy and guarantee the "right to be forgotten".
Machine unlearning due to DB deletes is qualitatively and quantitatively different from the setup studied in the ML literature, as (i) deletes are very commonplace, sometimes even more frequent than queries themselves, and (ii) typically, ``downstream tasks'' are different (e.g., AQP, SE, DG, DC, etc.). 
Nonetheless, the aims are the same: to update a trained model in such a way that the ``effect of deleted data is removed from the trained model'', while, at the same time, the model does not lose any knowledge it has learned about non-deleted data. 
Consider an AQP engine like DBEst++ \citep{ma2021learned} or a cardinality estimator like Naru/NeuroCard \citep{yang2019deep, yang2020neurocard}. These models essentially learn the data's underlying probability distributions and perform query inference based on these. When a cohort of data is removed, the models should be updated to reflect the correct densities (and/or correlations) for accurate inference. That is, we need to ensure that the updated model makes correct predictions when querying either the deleted rows and/or the remaining rows. 


Albeit an important problem, removing information from a trained neural network (without damaging the accuracy of the remaining data) is unfortunately a very challenging task. Ideally, one could remove the to-be-forgotten data from the DB and retrain a new model from scratch. However, as also has been shown by \citep{kurmanji2022detect}, training neural networks is prohibitively expensive.  

Furthermore, it is not easy to measure how well a model has truly forgotten the deleted cohort of data after an unlearning (forgetting) algorithm is applied, as defining an appropriate set of evaluation metrics is an open problem in and of itself. A contribution of our work is to employ metrics for unlearning that are appropriate in DBs,  drawing where possible inspiration from the ML literature that queries various aspects of a model's outputs like the error (accuracy) for classification tasks, the loss values, or the entropy of its predictions. 
We will elaborate on this in Section~\ref{sec:evaluation}.


\section{Machine Unlearning in Learned DB}

Our study aims to be general enough so that its conclusions are pertinent to different NNs and to different DB downstream tasks for which these NNs were designed.
(The selected NNs are not meant to imply any judgment on their being the best for selected tasks. Rather they are meant to boost the generality of drawn conclusions).

Consider a database relation $R$ with a set of columns \(\{C_1, C_2, ..., C_m\}\), with tuples $\{<c_1^i, c_2^i, \dots , c_m^i>\}_{i=1}^{|R|}$. This can be a raw table or the result of a join query. Also, let $f(\cdot;\theta)$ be a neural network with a set of trainable parameters $\theta$ that parameterize a function $f$. $f$ then could be used for any downstream task like AQP, SE, or DG, etc. Since here we wish to be general enough to handle different tasks/applications and different types of NNs,  $f$ might be trained with different objectives. 

Assume we have a dataset $D=\{<c_1^i, c_2^i, \dots , c_m^i>\}_{i=1}^{|R|}$ of tabular data with $|R|$ rows with $m$ columns. We will denote by $\theta^o$ the parameters of the (``original'', i.e. before unlearning) model trained on $D$, using a stochastic algorithm $\mathcal{A}$ like Stochastic Gradient Descent. That is, $\theta^o = \mathcal{A}(D, \theta^r)$, with $\theta^r$ denoting the random weights that the network is initialized with.
Now, at unlearning time, assume we are given a partition of $D$ into two disjoint sets of data rows: the ``delete-set'' (a.k.a ``forget-set'' or ``deleted data''), $D_d$, and the ``retain-set'', $D_r$, where $D_r = D \setminus D_d$; that is, the retain-set is the complement of the delete-set in $D$. Informally, the goal of machine unlearning is to transform the model's parameters into a new set of parameters $\theta^u = \mathcal{U}(D_r, D_f, \theta^o)$ that do not contain any ``knowledge'' about $D_d$ but retains all ``knowledge'' about $D_r$, where $\mathcal{U}$ denotes an unlearning algorithm that has access to the original parameters and the retain and forget sets. 

Formally defining unlearning and quantifying success is a challenging problem in and of itself \citep{thudi2022necessity}. A common viewpoint for the goal of unlearning in the literature is that we desire the weights $\theta^u$ to be indistinguishable from the oracle weights $\theta^*$ \citep{golatkar2020eternal}, or the outputs $f(\cdot; \theta^u)$ to be indistinguishable from the outputs $f(\cdot; \theta^*)$ \citep{golatkar2020forgetting}, where $\theta^* = \mathcal{A}(D_r, \theta^r)$ denotes the weights of the ``oracle'' of retraining from scratch using only $D_r$. More precisely, previous work defines the goal of (exact) unlearning as achieving $\mathbb{P}(\theta^u) = \mathbb{P}(\theta^*)$, with the probability distribution here being over model weights resulting from training with different random seeds (i.e. starting from different initial random weights $\theta^r$ and seeing a different ordering of mini-batches) or as achieving $\mathbb{P}(f(.;\theta^u)) = \mathbb{P}(f(.;\theta^*))$. For deep neural networks, the only known family of exact unlearning is based on retraining from scratch. To mitigate its inefficiency, recent work turned to approximate unlearning: achieving $\mathbb{P}(\theta^u) \approx \mathbb{P}(\theta^*)$ or $\mathbb{P}(f(.;\theta^u)) \approx \mathbb{P}(f(.;\theta^*))$, sometimes accompanied by theoretical guarantees about the quality of that approximation. 

In this work, we translate these definitions into a set of metrics that is suitable for DB tasks. Specifically, in learned DBs, generative models are frequently derived, that model the distribution in $D$, and subsequently we are interested in performance in downstream tasks. To be comprehensive in our study, 
we thus create two sets of metrics for quantifying unlearning quality: 1) with respect to the ``internal state'' of the model (i.e. its estimate of the probability distribution) and 2) with respect to the performance on downstream tasks. 
For completeness, in addition, we will also study discriminative models using a NN for an upstream DC task.
To quantify the success of unlearning in terms of downstream task performance, we utilize task-specific errors, where lower is better. Intuitively, we want the model to always make correct predictions (i.e. predictions matching the ground truth) for any query, whether the query relates to retained or deleted data.

\section{Literature Review}
\subsection{Machine Learning for Databases} \label{sec:ml_for_db}
Many ML-based DB components have emerged recently, using different models for different applications like Database indices 
\citep{kraska2018case,ding2020alex,nathan2020learning,ding2020tsunami}, 
learned cardinality/selectivity estimators 
\citep{hasan2020deep, yang2019deep, yang2020neurocard, zhu2020flat, wang2020we,factorjoin,negi2021flow}
and approximate query processors \citep{ma2021learned, ma2019dbest, thirumuruganathan2020approximate}, 
query optimization and join ordering \citep{marcus2019neo, kipf2018learned}, cost estimation  \citep{zhi2021efficient, siddiqui2020cost}, and auto-tuning databases \citep{van2017automatic,li2019qtune,zhang2019end}. 
However, these do not provide any insights or solutions to adapt to changes due to DB data updates.
The recent works in \citep{li2022warper,kurmanji2022detect} tackled distribution changes for CE tasks and CE/AQP/DG tasks respectively. ML methods have been used also for deriving samples of join query results without executing the join \citep{shanghooshabad2021pgmjoins}.
However, these approaches cannot account for data deletions, which as we shall see pose different updatability problems to NN-based learned DBs. 

\Rc{As in \citep{kurmanji2022detect, li2022warper}, we also consider updates in batches - the isolated effect of single data rows in learned models is negligible. 
In environments like Online Transactional Processing (OLTP), however, update frequencies are higher and data may be deleted/inserted in single records. In such a setting the approach would be: The set of updates in a batch would be preprocessed and the final insert/delete records/operations would be identified. Delete records would comprise the forget set for unlearning. Insert records would comprise the new batch in \citep{kurmanji2022detect, li2022warper}). As the results in Section \ref{sec:experiments} show, unlearning can work well with smaller or larger batches and in conjunction with data insertions.}

\subsection{Machine Unlearning} \label{unlearning_literature}
The problem of machine unlearning was first introduced in \citep{cao2015towards}, where they provide an exact forgetting algorithm for statistical query learning. \citep{bourtoule2021machine} proposed a training framework by sharding data and creating multiple models, enabling exact unlearning of certain data partitions. \citep{ginart2019making} was the first paper to introduce a probabilistic definition for machine unlearning which was inspired by Differential Privacy \citep{dwork2014algorithmic}, and formed the origin of the idea of the model indistinguishability definition discussed above. More recently, \citep{guo2019certified,izzo2021approximate,sekhari2021remember, wu2020deltagrad} built upon this framework  
and introduced unlearning methods that can provide theoretical guarantees under certain assumptions. \citep{mahadevan2021certifiable} surveyed methods for linear classification, showing different certifiability versus efficiency trade-offs. 

Recently, approximate unlearning methods were developed that can be applied to deep neural networks. \citep{golatkar2020eternal} proposed an information-theoretic procedure for removing information about $D_d$ from the weights of a neural network and \citep{golatkar2020forgetting,golatkar2021mixed} proposed methods to approximate the weights that would have been obtained by unlearning via a linearization inspired by NTK theory \citep{jacot2018neural} in the first case, and based on Taylor expansions in the latter.

However, \citep{golatkar2020eternal,golatkar2020forgetting} scale poorly with the size of the training dataset, as computing the forgetting step scales quadratically with the number of training samples. \citep{golatkar2021mixed} addresses this issue, albeit under a different restrictive assumption. Specifically, they assume that a large dataset is available for pre-training that will remain ``static'', in the sense that no forgetting operations will be applied on it; an assumption that unfortunately can't always be made in practice.

Some recent works suggest modifying the original model's training for better unlearning in the future. \citep{thudi2022unrolling} propose a regularizer to reduce `verification error' making unlearning easier. However, it may impact model performance. \citep{zhang2022prompt} introduce a process with quantized gradients and randomized smoothing to avoid future unlearning, but large changes in data distribution, as a result of deletion, may exceed the `deletion budget' invalidating assumptions. More recent works try to directly identify the parameters in the original model that are significantly influenced by the forget-set, aiming to modify these parameters to eliminate the impact of the forget-set.
\citep{foster2023fast, shi2023deepclean} leverage fisher information scores to identify the important parameters for the forget-set. \citep{shi2023deepclean} take a straightforward approach by fine-tuning the model on the retain-set while keeping the remaining parameters frozen. In contrast, \citep{foster2023fast} tries to `dampen' those parameters while minimizing adverse effects on those essential to the retain-set. \citep{jia2023model} introduce a 'sparsity-aware' unlearning technique, integrating unlearning through fine-tuning on the retain-set with a sparsification policy employing model pruning techniques.  \citep{fan2023salun}, on the other hand, proposes a 'saliency-aware' unlearning approach, utilizing the loss function's gradient to learn a mask identifying 'salient' parameters related to the forget-set. These parameters are subsequently unlearned using existing unlearning baselines such as `random labelling'. Despite the diversity in these methods, a common challenge persists—efficiently identifying and modifying the important parameters tied to the forget-set without adversely affecting those essential to the retain-set.

On the other hand, SCRUB \citep{kurmanji2023towards} is a recent machine unlearning method for computer
vision image classification tasks that
scales better than previous works without making restrictive assumptions.
SCRUB reveals different requirements and metrics for different unlearning applications (e.g., removing biases, correcting erroneous mislabelling or attack-poisoned labels, and preserving user privacy).
SCRUB is shown to be the most consistently well-performing approach on a wide range of architectures, datasets and metrics.

\section{Learning Tasks (Applications)} \label{sec:applications}

We will study unlearning in the context of four well-studied, key tasks for learned DBs and data analytics (AQP, SE, DG, DC), using each a different NN type (as used in previous work). \Ra{We now give an overview of each of these tasks.}

\subsection{Downstream DB Applications}
\Ra{\textbf{Approximate Query Processing}. AQP approximates the results of aggregation 
queries.  
This is a key task in analytical DBs, particularly for very large 
tables, where obtaining exact results can be prohibitively expensive \citep{ma2021learned, thirumuruganathan2020approximate, hilprecht13deepdb, kurmanji2022detect}. }

\Ra{\textbf{Selectivity Estimation}. SE refers to the process of estimating the number of rows that a query will return. 
This is key for query optimization in RDBMSs, as it helps the query planner make informed decisions about best execution plans 
\citep{yang2019deep, yang2020neurocard, kurmanji2022detect}.}

\Ra{\textbf{Data Generation}. (Synthetic) data generation involves creating artificial data that mimics the characteristics of real-world data. 
DG is important for 
overcoming issues of insufficient or sensitive real data,
thus dealing with privacy concerns or data scarcity issues.} 

\Ra{\textbf{Data Classification}. DC has numerous real-world applications. 
 The integration of classifiers within DBMSs is key in enhancing business intelligence and analytical services. 
 For example, Microsoft SQL Server provides a `data discovery and classification' feature. Similarly, Google's BigQuery empowers data mining processes through a repertoire of classifiers. Such classifiers typically use a categorical attribute whose values define the different classes according to which tuples are classified.}

\subsection{Machine Learning Models}

\textbf{Mixture Density Networks (MDNs) for AQP}. MDNs consist of an NN to learn feature vectors and a mixture model to learn the \textit{probability density function} (pdf) of data. \citet{ma2021learned} propose DBest++ which uses MDNs with Gaussian nodes to perform AQP. For the Gaussian Mixture, the last layer of MDN consists of three sets of nodes \(\{\omega_i, \mu_i, \sigma_i\}_{i=1}^M\) that form the pdf according to Eq. \ref{mdneq}, where \(M\) is the number of Gaussian components in the mixture. 

Let \(y\) be the dependent variable (a target numerical attribute), $\mathbf{x}$ the vector \((x_1, ..., x_n)\) of independent variables (a set of categorical attributes). We also define $f$ to be a NN that takes the encoded input vectors $\mathbf{x}$ and transforms them into learned ``feature vectors'' $h$. The likelihood under the mixture of Gaussians is then given by:


\begin{equation}\label{mdneq}
\hat{P}(y| h) = \sum_{i=1}^{M}\omega_i\mathscr{N}( h ; \mu_i, \sigma_i)
\end{equation}
where
\begin{equation}
\omega = g_1(h ; \theta^{\omega}); \mu = g_2(h ; \theta^{\mu}); \sigma = g_3(h ; \theta^{\sigma}); h = f(\mathbf{x} ; \theta^{base})
\end{equation}

where each of $g_1$, $g_2$ and $g_3$ is a single-layer network that produces the weight (``mixing proportion'') \(\omega_i\), the mean \(\mu_i\) and standard deviation \(\sigma_i\), respectively, for each of the \(i\) Gaussians in the mixture.
Since the mixing proportions should sum up to 1, $g_1$ has a Softmax activation function, whereas $g_2$ and $g_3$ use a ReLU activation. 
We train all parameters of the system, ($\theta^{base}$, $\theta^{\omega}$, $\theta^{\mu}$, $\theta^{\sigma}$)  
jointly, by minimizing the negative log of the likelihood in Eq. \ref{mdneq}. 


\textbf{Deep Autoregressive Networks for SE}. The Naru and NeuroCard cardinality/selectivity estimators \citep{yang2019deep, yang2020neurocard} use deep autoregressive networks (DARNs) to approximate a fully factorized data density. DARNs are generative models capable of learning full conditional probabilities of a sequence using a masked autoencoder via maximum likelihood. Once the conditionals are available, the joint data distribution can be represented by the product rule as follows:
\[
\hat{P}(x_1, x_2, \dots, x_m) = \hat{P}(x_1)\hat{P}(x_2|x_1)\dots \hat{P}(x_m|x1,\dots ,x_{m-1})
\]

where \(x_i\) is an attribute in a relation \(R\) with \(m\) columns and the probability of the \(i^{th}\) conditional, $\hat{P}(x_i|x1,\dots ,x_{i-1})$, is parameterized via a neural network $f(\cdot ; \theta_i)$. Naru and NeuroCard use cross-entropy between input and conditionals as the loss function, to train the parameters of the NNs $\theta_1, \dots, \theta_n$. 

\textbf{Variational Autoencoders for DG}. VAEs \citep{kingma2013auto} are a commonly-used model for data generation. A VAE is a probabilistic auto-encoder, that uses a pair of neural networks to parameterize an encoder and a decoder module. In DB systems,  \citep{thirumuruganathan2020approximate} used VAEs to build AQP engines, \citep{hasan2020deep} exploited them for CE, and \citep{xu2019modeling} introduced a synthetic tabular data generator called \textbf{Tabular-VAE} (TVAE). VAEs are trained using a different loss function, known as Evidence-Lower-Bound (ELBO) loss (which amounts to a lower-bound estimation of the likelihood). Here we will use \textbf{TVAE} for learned synthetic tabular data generation, which is of particular importance in privacy-sensitive environments, or when data is scarce for data augmentation purposes, or when wishing to train models over tables and accessing raw data is costly.

{\textbf{Deep NNs for Classification for DC}. Our previous three applications are based on {\it generative models}, trained in an unsupervised/semi-supervised fashion. To complete the picture, we add a data classification (DC) task, using a {\it discriminative} NN model. 
Note that there is no downstream task.
Traditionally, DC over tabular data has been tackled using decision trees, random forests, support vector machines, etc. 
However, deep NNs have emerged as a powerful alternative that can effectively learn complex relationships and patterns in tabular data. 
Deep NNs for tabular DC leverage various architectures (e.g., feedforward NNs, convolutional NNs (CNNs), or recurrent NNs (RNNs)). 
\citet{gorishniy2021revisiting} have shown that, for tabular data, ResNet-like architectures are strong performers. 
Residual Neural Network (ResNet) \citep{jian2016deep}, is a family of deep learning architectures that have significantly advanced the field. 

\section{Unlearning Methods} \label{sec:baselines}
In this section, we describe key baselines and a state-of-the-art method that has been proposed by the unlearning research in the ML community, in the context of classification tasks in computer vision. Each of these offers a different procedure to utilize the retain-set $D_r$ and / or the delete-set $D_d$ to achieve unlearning.

\textbf{Retrain}. The ``oracle'' solution is to remove the to-be-forgotten data from the DB and retrain a new model on only $D_r$. Retraining neural networks ``from scratch'', however, is prohibitively expensive and not practical in many settings \citep{kurmanji2022detect}. 

\textbf{Stale}. 
This leaves the model stale as it was trained on the original data $D$. 
For DB tasks this is more complex as most of the learned DB components comprise learned NNs, and non-learned modules such as the auxiliary meta-data. The stale baseline here updates the meta-data (like table cardinalities, or frequency tables), and leaves the learned model stale. 

\textbf{Fine-tune}. This simple baseline fine-tunes the original model for a small number of epochs on the remaining rows. Concretely, it continues training the NN with data from $D_r$ only, steering the model towards forgetting the delete-set. 
In ML classification tasks, it has been shown that this method will not erase the information completely \citep{golatkar2020eternal}. 
Interestingly, fine-tuning has also been suggested by some of the learned DB components to support insertion updates \citep{yang2019deep, ma2021learned}. 
There, fine-tuning is performed on the new data to force the model to learn it. 
This method has been shown not to perform well with OOD data insertions \citep{kurmanji2022detect}. 

\textbf{NegGrad}. 
An interesting idea 
is to continue training the model, but instead of using gradient descent on only $D_r$ as in \texttt{Fine-tune}, use gradient \textit{ascent} on only $D_d$ (or equivalently, gradient descent on $D_d$ with a negated gradient; earning the nickname NegGrad). The intuition for this is to attempt to ``delete'' $D_d$ by maximizing the loss on that data, aiming to ``undo'' the process that the network had previously undergone to learn that data. 

\textbf{NegGrad+}. 
NegGrad may degrade the performance on the retain set, since it has no incentive to protect useful information from being deleted when performing the gradient ascent. Intuitively, if retained data is ``similar'' to data in the delete-set, then \texttt{NegGrad}'s objective of maximizing the loss on $D_d$ may indirectly also lead to maximizing the loss on (parts of) $D_r$, which is of course undesirable. To address this, we  use a stronger baseline  
that simultaneously performs gradient ascent (as in \texttt{NegGrad}) on the delete-set and gradient descent (as in \texttt{Fine-tune}) on the retain-set, aiming to strike a good balance between maximizing the loss on $D_d$ but keeping it small on $D_r$. We refer to this stronger baseline as \texttt{NegGrad+}. Formally, it obtains the unlearned weights $\theta^u$ by initializing them from $\theta^o$ and then minimizing the following  w.r.t $\theta^u$:

\begin{equation}\label{eq:neggrad_loss}
   \beta \displaystyle\frac{1}{|D_r|} \displaystyle\sum_{x^r \in D_r} l(f(x^r;\theta^u)) - (1-\beta) \displaystyle\frac{1}{|D_f|} \displaystyle\sum_{x^d \in D_d} l(f(x^d;\theta^u))
\end{equation}
Where $l$ is a task-specific loss function like negative likelihood and $\beta \in [0,1]$ is a hyperparameter. Note that we can recover Fine-tune by setting $\beta$ to 1 and NegGrad by setting $\beta$ to 0.

\textbf{SCRUB}. This is a more sophisticated state-of-the-art algorithm for unlearning in image classification \citep{kurmanji2023towards}. SCRUB makes use of a teacher-student framework where the teacher is the original model, trained on the entire dataset, and the student is initialized from the teacher and is subsequently trained to keep only the information from the teacher that pertains to the retain set. Specifically, the student is trained with two objectives: one that minimizes the distance between the teacher and the student for the retain-set, and one that maximizes this distance for the delete-set, leading to surgically removing information about the delete-set. Similarly to NegGrad+, SCRUB aims to find a sweet spot between deleting information about $D_d$ while retaining information about $D_r$ but does in the form of (positive and negative, for $D_r$ and $D_d$, respectively) knowledge distillation from the ``all-knowing'' original model. Formally, SCRUB obtains the unlearned (student) weights $\theta^u$ by initializing them from the original (teacher) weights $\theta^o$ and then minimizing the following objective w.r.t $\theta^u$:

\begin{equation}\label{eq:scrub_se}
\beta \displaystyle\frac{1}{|D_r|} \displaystyle\sum_{x^r \in D_r} d(x^r; \theta^o,\theta^u)
   - (1-\beta) \displaystyle\frac{1}{|D_f|} \displaystyle\sum_{x^d \in D_d} d(x^d; \theta^o,\theta^u)
\end{equation}
where $d$ is a measure of the distance between the student and teacher models, to be defined in an application-dependent manner. While SCRUB is originally defined for classification, using KL-divergence for $d$, here we propose an adaptation of it for two of our DB tasks: SE and AQP. 

For SE, where we use deep autoregressive models to estimate the joint (over columns) probability density $\hat{P}$ of the data, we define $d$ as the KL divergence between the estimated probabilities of the teacher and student. Formally, 
\begin{equation}
d(x; \theta^o,\theta^u) = KL( \hat{P}(x; \theta^o) || \hat{P}(x; \theta^u))
\end{equation}

For AQP, where Mixture Density Networks are used, it is not straightforward how to apply SCRUB.
We propose the following form for $d$ to capture the discrepancy between the mixture distributions of the teacher and student: 
\begin{equation}
d(x; \theta^o,\theta^u) = \frac{1}{M} \sum_{i}^M (MSE(\mu_i^o, \mu_i^u) + MSE(\sigma_i^o, \sigma_i^u) + KL(\omega_i^o || \omega_i^u))
\end{equation}
where for a given $x$, $\omega$ is obtained as $g_1(f(x; \theta^{base}); \theta^{\omega})$, and analogously for $\mu$ and $\sigma$, as explained for MDNs in Section~\ref{sec:applications}. $MSE$ is Mean Squared Error, M is the number of components in the mixture, and the parameters $\theta$ (each of $\theta^u$, $\theta^o$, correspondingly) are the set:
\begin{equation}
 \theta = \{ \theta^{base}, \theta^{\omega}, \theta^{\mu} , \theta^{\sigma} \}
\end{equation}



\Rb{\textbf{SISA}. Sharded, Isolated, Sliced, and Aggregated (SISA) is an ensemble method that first splits the data into  \(p\) disjoint partitions, and further slices each partition into \(s\) slices. SISA trains a constituent model for each partition, by incrementally incorporating slices of that partition.
During unlearning, when a delete-set is requested to be unlearned, SISA finds all the models that have been trained with the examples from the delete-set, removes those examples from the dataset and retrains the affected models from scratch. As such, SISA is an `exact unlearning' method, like Retrain, unlike the rest of the baselines we consider which perform `approximate unlearning'.}

\Rb{Since SISA was originally built for classification tasks, it uses a majority vote aggregation. However, for most of the applications that we study, including AQP, SE and DG this aggregation is not applicable. Therefore, we design new aggregations: For AQP and SE where the models evaluate a workload of \texttt{sum} or \texttt{count} queries, we calculate the result of the query using each model and sum up the results. For DG, we generate samples using each model, and concatenate all the samples to form the final synthetic data.}


\section{Measuring the Impact of Deletion}\label{sec:evaluation}
As mentioned earlier, this is an open problem of its own right in the ML literature. And for learned DBs, as we discuss below, several adjustments need to be made. 

In the ML literature unlearning is mostly studied for classification problems, and the accuracy of the retain-set and the test-set are used to measure the model's performance on the remaining data and its generalization, respectively. 
We desire an unlearning algorithm that successfully `forgets' the delete-set without deteriorating either of these. 
We measure the `forget quality' using the accuracy on the delete-set which is ideally close to the delete-set accuracy of the \texttt{Retrain} oracle (that truly never trained on the delete-set). 

The commonly-studied applications for unlearning in the ML literature (i.e., classification) concern the direct output of learned models (i.e., upstream tasks). 
Our setting is different; our learned DB models are developed for \textit{downstream tasks}, that are not the immediate output of the trained model. 
For instance, the AQP engine in \citep{ma2021learned} infers the query answer using an integral estimation over the learned Gaussians produced by MDNs. 
Similarly, the cardinality estimators in \citep{yang2019deep,yang2020neurocard} perform a progressive sampling over the learned DARN to infer the cardinality. 
Given this, we will care both about what the model has learned and unlearned, as well as the accuracy of the downstream tasks.
We establish a distinction between, on the one hand, evaluating the quality of unlearning with respect to the internal state of the model itself (how well has it forgotten the requested data) and, on the other hand, how its performance on the downstream tasks of interest is affected. This distinction is interesting from a scientific perspective, as it enables the study of several research questions concerning unlearning in generative pretrained models: Can we achieve the desired outcome on a set of downstream tasks without having optimally amended the model's internal state in light of deletions? What is the relationship between unlearning quality upstream and downstream? In this work, we initiate this investigation and propose a set of metrics for each of these evaluation facets.

To study the downstream task's performance, we use the original metrics that have been used to evaluate the task, with a small change. Inspired by the accuracy evaluation in unlearning in ML, we divide the evaluation workload into two separate groups that target the delete-set and the retain-set. More specifically, for AQP and CE, one workload only queries the deleted rows, and the other workload queries the remaining rows. In both cases, the lower the error, the better (unlike other ML applications where higher error is desirable / indicates better forgetting). 
Intuitively, considering a COUNT query for example, when the user requests a query that targets the deleted part, the engine should correctly answer 0. For DG, the generated data could be used for different tasks. We follow the evaluation in \citep{xu2019modeling} where we use a classification task and evaluate on the test-set and measure forgetting via accuracy on the delete-set.

Additionally, we evaluate unlearning in the model's internal state in two ways. First, by inspecting the likelihood, we can assess what data is ``likely'', or ``compatible'' with the model's internal understanding of the distribution. Indeed, likelihood is often the training objective of learned DB models and, in fact, in some cases, like in MDNs, the exact likelihood is available through the mixture of Gaussians. Intuitively, we would like the model to assign a higher likelihood to the retain-set on average, and a lower likelihood to the delete-set.
Second, since learned DB systems are usually generative models that learn the data density, we can directly inspect the learned probability distribution via sampling, before and after deletion. Intuitively, we want the unlearning process to modify the learned distribution in such a way that it accurately reflects the updated true ``state of the world'' after the deletions. 

\Rb{Finally, we use Membership Inference Attacks (MIAs) to asses the forgetability of the models. In an MIA, the adversary tries to infer whether a data point has been involved in training a model. We focus on the common black-box attack setting where the attacker observes only outputs of the model. Designing MIAs generally, and especially as a metric for forgetability, is an open research problem. Nevertheless, we take two attacks that have been used in the unlearning literature and apply them to our DC models.}

\section{Experimental Evaluation}\label{sec:experiments}

We empirically evaluate and analyze the aforementioned unlearning algorithms, in the context of several applications and evaluation metrics. We consider two scenarios: Deletion in ``one-go'', where a single round of deletion is performed, as well as ``sequential deletion'', where several deletion requests are carried out sequentially. The code and information for reproducibility and availability purposes can be found at \url{https://github.com/meghdadk/DB_unlearning.git}.

\subsection{Experimental Setup}

\subsubsection{Datasets}
We have used three real-world tabular datasets, typically found in the literature for our downstream tasks, namely: Census: 48k rows, Forest: 580K rows, and DMV: 11M rows. 

\subsubsection{Delete Operations}
Deletes have the following skeletons:

\begin{lstlisting}[mathescape=true,
    basicstyle=\footnotesize, %or \small or \footnotesize etc.
]
1. DELETE FROM $Table$ WHERE L <= $att$ <= U
2. DELETE FROM $Table$ WHERE (L <= $att$ <= U) and 
                           ($row\mhyphen index$ % 2 == 0)
\end{lstlisting}
where $att$ is a numerical attribute, $L$ and $U$ define the range of values that will be deleted, and $row\mhyphen index$ is a primary incremental index started from 1. In the former, a whole data subspace is deleted (``full deletion'') whereas in the latter, only some parts within a subspace are deleted (``selective deletion'').


For Census, $att$ refers to the $age$ attribute, where $L=30$ and $U=35$. 
For Forest, $att$ refers to $Elevation$ and  $L=2500$, $U=2700$. 
For DMV, $att$ is $max\mhyphen gross\mhyphen weight$, $L=7000$, and $U=8200$. 
In all cases, full (selective) deletion queries delete ~10\% (~5\%) of the dataset.

\subsubsection{Queries and Metrics} 
For the AQP and SE tasks, after a delete is performed, two types of queries are generated: Query-Retain (QR) targets only the remaining (non-deleted) data. 
On the other hand, Query-Delete (QD) queries only target deleted data. 
For QR, we report the ``relative error'' (relative to the ground truth) to evaluate AQP and SE tasks. For QD, however, the ground truth is always 0 for both tasks (since the sum, mean and count over deleted rows are 0); thus, we report the ``absolute error'' instead of the relative error, to avoid a division by 0.
Furthermore, for DBEst++ MDNs, as their mixture of Gaussians provides the exact likelihood, we also report the average likelihood for QD and QR. 

For the DG task we evaluate the model's data generation quality via the accuracy of an XGboost classifier trained on the synthetic samples generated by the model (TVAE) after training, as in \citep{xu2019modeling}. 
We hold out 30\% of the synthetic table as the test set and train a classifier with XGBoost. Then we predict the classes of the held-out data test set. We report the \textit{macro f1-score} for the classifier. For Census, Forest and DMV, we use: \textit{income}, \textit{cover-type}, and \textit{fuel-type}, as the target class, respectively.
Here we created a smaller version of DMV with only 1M records, as training TVAE on the whole DMV is very time/resource-consuming. For this smaller DMV, instead of forming the deletion query via a range, as shown above, we delete all rows that satisfy \(State= `OK\textrm'\).

For the DC task, we use: \textit{marital-status}, \textit{cover-type}, and \textit{fuel-type} for Census, Forest and DMV, as the target class, respectively. 
We split the tables into a 80\%-10\%-10\% splits of train-validation-test. 

\subsubsection{Workloads} 
Each model is evaluated using 2000 randomly generated QR and 2000 QD queries. 
For Naru/NeuroCard, we use their generator to synthesize QR and QD queries: It randomly selects the number of filters per query. For Forest, this number is randomly selected from the range [3,8], for Census from [5,12], and for DMV from [5,12] 
Then, it uniformly selects a row of the table and randomly assigns operators \([=,>=,<=]\) to the columns corresponding to the selected filters. 
Columns with a domain less than 10 are considered categorical and only equality filters are used. 

For DBest++, we randomly select a \(lower\mhyphen bound\) and a \(higher\mhyphen bound\) for the range filter and uniformly select a category from the categorical column for the equality filter.

\begin{lstlisting}[mathescape=true,
    basicstyle=\footnotesize, %or \small or \footnotesize etc.
]
SELECT AGG($\cdot$) FROM $Table$ WHERE $F_k$ AND $F_{n}$
\end{lstlisting}


where AGG is an aggregation function over a numerical attribute, and \(F_k\) ( \(F_n\) ) is a filter over a categorical (numerical) attribute.
Specifically, \(F_n\) is a range operation and the aggregation function is \texttt{COUNT}, \texttt{SUM}, or \texttt{AVG}. 
We select the following columns from each dataset: Census:[\texttt{age, country}]; Forest:[\texttt{slope, elevation}]; DMV:[\texttt{body type, max gross weight}]; where the first/second attribute is categorical/numeric. Naru/NeuroCard is a cardinality estimator and we only evaluate \texttt{COUNT} queries for it. 

\subsubsection{Models' configurations}
For DBEst++, for Census, we build an MDN with 2 fully connected layers of size 128 each, and use a MoG of size 30 for the last layer. For Forest and DMV, we use the same number of fully connected layers with the same size, but with 80 MoG for the last layer. We train the \texttt{Original} and \texttt{Retrain} models for 50 epochs, with a learning rate (lr) of 0.001 (decaying with a rate of 0.1 at epochs 10, 20, 30), and a batch size (bs) of 128, using the Adam optimizer. For Naru/NeuroCard, and TVAE, we use the same configuration as in the original work and tune hyper-parameters for smaller errors, for the datasets that have not been used in the original work. For the unlearning baselines, we tune all the hyper-parameters of the models including lr, bs, decay rate, optimizer, and the method's specific hyper-parameters, to achieve the smallest retain error and forget error.

\Rb{For SISA, we set \(p=10\) and \(s=10\) for the DC task, and \(p=5\) and \(s=5\) for other applications. For DC, we use a smaller neural network with 3 fully connected layers. For AQP/DBEst++ application we use an MDN with fully connected layers of size 64 and 15 MoG for Census and 30 for Forest and DMV. For SE/Naru-NeuroCard and DG/TVAE we decrease the models' depth by 1 layer for each of the encoder and the decoder. Finally, for better efficiency of SISA, we sort each dataset based on the column that is queried for deletion.}

\subsection{Deletion in ``One-Go'' (Large Batch)}
Deletion (in ``one-go''' -- i.e. using a large batch of deletions) is studied using both downstream task and model-internal metrics.

\subsubsection{Results on Downstream Tasks}
\textbf{AQP/DBEst++}.  
These results are reported in Tables \ref{tab:mdn_full_intervals} and \ref{tab:mdn_selective_intervals}.
Overall, the main conclusions from \autoref{tab:mdn_full_intervals} are as follows:
We first notice that \texttt{NegGrad} performs well for QD but poorly for QR, which is expected as it has no incentive to retain knowledge of the retain-set, only to erase knowledge for the delete-set.
For QR-count, \texttt{Retrain}, \texttt{Fine-tune}, and \texttt{SCRUB}, all perform similarly (with mostly overlapping CIs) 
\texttt{NegGrad+} follows, with a slightly worse performance for Census and Forest, but strong performance on DMV.
For QR-sum, the findings are very similar.

For QD, as expected, \texttt{NegGrad} performs very well, as it was designed to update the model specifically for deletes. 
With respect to \texttt{Stale}, we would expect it to have very poor performance for deleted data, as this has not been updated to learn it.
This is indeed the case.
For QD-count: We note that \texttt{NegGrad+} is a top performer, hand-in-hand with \texttt{Retrain} and that 
\texttt{Fine-tune} does very well (very close to \texttt{Retrain}).
\texttt{SCRUB} follows in performance, but in absolute terms, it achieves small errors (e.g., being much closer to \texttt{Retrain} than to \texttt{Stale}. And for DMV \texttt{SCRUB}'s performance has overlapping CIs with \texttt{Retrain}.
For QD-sum, the conclusions are very similar to those of QD-count.
\texttt{NegGrad} may be a top performer for QD, but note that its performance for QR is dismal.

For selective deletion (\autoref{tab:mdn_selective_intervals}), perhaps surprisingly, \texttt{Stale} appears to be doing better than would be expected.
However, recall that: (i) queries here cover both remaining and deleted data simultaneously, due to the interleaved manner with which the retain and forget sets are defined, and \texttt{Stale} is expected to do well for retained data, and (ii) \texttt{Stale} may be leaving the model unchanged but it does update other statistics (such as frequency tables) used in the end to predict the aggregate. Given these facts, the good performance on this interleaved set is not so surprising. 
This finding highlights why we used downstream task errors $and$ internal model metrics. Specifically, looking at  \autoref{tab:mdn_selective_intervals}
we clearly see that \texttt{Stale} has a much higher likelihood for deleted data than the other unlearned models. This is so because \texttt{Stale} has not unlearned and deleted data is still part of its training set. We will discuss this mismatch between internal state and downstream task metrics in later sections.

\textbf{SE/Naru-NeuroCard}. 
We have reported the results in \autoref{tab:naru_grouped_ci}.
The results are for both full and selective deletion. 
As mentioned earlier, Naru-NeuroCard involve a sampling process during inference that automatically zeroes-out the queries on the delete-set as the sample does not find any record from that region of the table to sample from. 
We observe that \texttt{Fine-tune} is the top performer. 
In fact, fine-tuning the model results in better accuracy than retraining from scratch. This is an interesting finding of its own right, as sometimes fine-tuning may reinforce old knowledge while unlearning as well.  
Similarly, \texttt{NegGrad+} also performs great.
\texttt{SCRUB} is also a top performer, except for full deletion on DMV.
\texttt{Stale} also does well (as also observed above for  AQP). But, note that \texttt{Stale} performs poorly for full and selective deletion for DMV. And as before, its performance for full deletion is worse than that for selective deletion.
Again, it is crucial to evaluate across models, downstream tasks, internal model metrics, and datasets for a complete picture.
Finally, as before, \texttt{NegGrad} degrades the model drastically. 

\textbf{DG/TVAE}. Finally, in \autoref{tab:tvae_grouped} we show the results stemming from the classification accuracy over the synthetic data generated by TVAE across different unlearning methods and datasets. 
We do not report \texttt{SCRUB} as it is unknown how to apply its loss and integrate it with a TVAE.
We also do not show numbers for \texttt{NegGrad} as its performance was very poor, suffering from exploding gradients, even with very small learning rates and a very small number of iterations.
The first main observation is that all methods perform very close to each other. Even \texttt{Stale}. This experiment helps bring to the surface additional interesting issues. Why do $all$ methods, even \texttt{Stale}, perform equally? This occurs because the delete operation in this particular case happens to not affect the classification task. We investigated this by computing the Pearson correlation coefficient values among the dependent and independent variables (omitted for space reasons) and we found that essentially all correlations remain unaffected before and after deletion.
Please note that this reflects a perfectly reasonable real-world scenario as not all delete operations affect downstream task accuracy.
Nonetheless, an additional concern for such cases is whether the ML models themselves were affected at all by the deletion.
And this is why model-internal metrics should be used to reveal such differences. 
In fact, as we shall show in the next subsection, there are clear differences between models of different methods with respect to the effect of even such deletion operations; and, it is shown that the \texttt{Stale} model fails to register the deletion. 

\begin{table*}[]
\small
  \centering
    \caption{Full deletion results for AQP/DBEst++. For QR we show the average relative-error and 95 \% CI. For QD we show the average absolute-error and 95 \% CI. ``count'' (``sum'') refer to queries with COUNT (SUM) aggregation function. Likelihood numbers show the average likelihood the MDN model predicts for the retained and the deleted rows using Eq. \ref{mdneq}.}
    \vspace{-0.35cm}
    \label{tab:mdn_full_intervals}
    
    \begin{tabular}{c | c | c c | c c | c c }
\toprule
\multirow{2}{*}{dataset}&\multirow{2}{*}{model}&\multicolumn{2}{c|}{Query-Retain - QR}&\multicolumn{2}{c|}{Query-Delete - QD}&\multicolumn{2}{c}{likelihoods}\\
&&count&sum&count&sum&remain&delete  \\
\midrule
\multirow{6}{*}{Census}
&Retrain & 16.26±1.06 & 16.53±1.11 & 0.09±0.04 & 2.29±0.83 & 0.88 & 0.04\\
&Stale & 22.44±1.38 & 23.03±1.54 & 26.62±6.60 & 824.70±211.52 & 0.74 & 0.98\\
&Fine-tune & 16.88±1.10 & 17.58±1.39 & 0.83±0.19 & 24.69±5.15 & 0.86 & 0.13\\
&NegGrad+ & 20.20±1.39 & 19.46±1.46 & 0.27±0.02 & 11.34±6.81 & 0.87 & 0.07\\
&NegGrad & 153.27±15.78 & 141.54±13.31 & 0.00±0.00 & 0.00±0.00  & 0.22 & 0.00 \\
&SCRUB & 17.39±1.14 & 17.83±1.35 & 1.94±0.51 & 57.48±14.23 & 0.85 & 0.18\\

\midrule
\multirow{6}{*}{Forest}
&Retrain & 9.50±1.02 & 9.25±0.84 & 0.00±0.00 & 0.69±0.13 & 1.36 & 0.01\\
&Stale & 48.12±34.32 & 24.59±3.37 & 4.23±0.22 & 9148.42±512.56 & 1.24 & 0.49\\
&Fine-tune & 10.15±1.08 & 9.96±0.98 & 0.01±0.00 & 14.68±5.97 & 1.36 & 0.01\\
&NegGrad+ & 13.55±1.10 & 13.81±1.07 & 0.00±0.00 & 1.75±3.20  & 1.36 & 0.00 \\
&NegGrad & 137.93±37.00 & 157.15±62.51 & 0.00±0.00 & 0.63±0.88 & 1.01 & 0.00 \\
&SCRUB & 11.16±1.47 & 10.29±0.93 & 0.05±0.00 & 72.93±7.49 & 1.36 & 0.03  \\

\midrule
\multirow{6}{*}{DMV}
&Retrain & 85.06±35.36 & 82.96±21.45 & 0.46±0.17 & 2448.82±711.35 & 85.09 & 8.27\\
&Stale & 56.05±11.46 & 58.42±7.77 & 2.13±0.70 & 13367.03±2780.23 & 40.78 & 42.90\\
&Fine-tune & 106.07±49.79 & 89.86±26.12 & 1.10±0.39 & 5545.70±1251.51 & 39.68 & 16.42\\
&NegGrad+ & 81.52±25.01 & 119.98±91.18 & 0.52±0.15 & 3457.04±1071.17 & 41.65 & 4.58\\
&NegGrad & 8083.59±527.80 & 1131923741.20±64946.12 & 0.00±0.00 & 0.00±0.00 & 2.78&0.00 \\
&SCRUB & 69.49±14.05 & 102.54±70.92 & 0.50±0.25 & 2642.01±487.11 & 46.06 & 12.77\\

\midrule

    \end{tabular}
\end{table*}
\begin{table}
\small
  \centering
    \caption{Selective deletion results for AQP/DBEst++. In this setting queries cover ranges of remained and deleted values and the ground truth is therefore never zero. So, we show the average relative-error and 95 \% CI. The likelihood numbers show the average likelihood the model predicts for the retained rows and the deleted rows using Eq. \ref{mdneq}.}
    \vspace{-0.35cm}
    \label{tab:mdn_selective_intervals}
    \resizebox{.95\linewidth}{!}{
    \begin{tabular}{c | c | c c | c c}
\toprule
\multirow{2}{*}{dataset}&\multirow{2}{*}{model}&\multicolumn{2}{c|}{Queries}&\multicolumn{2}{c}{likelihoods} \\
&&count&sum&remain&delete \\
\midrule
\multirow{6}{*}{Census}
&Retrain & 15.36±1.09 & 14.12±0.97 & 0.78 & 0.61 \\
&Stale & 17.68±1.21 & 16.15±0.99 & 0.77 & 0.98\\
&Fine-tune & 16.02±1.08 & 14.69±0.97 & 0.78 & 0.64\\
&NegGrad+ & 15.31±1.01 & 14.16±0.89 & 0.78 & 0.59 \\
&NegGrad & 166.76±19.62 & 157.35±18.32 & 0.19 & 0.00\\
&SCRUB & 15.66±1.07 & 14.48±0.87 & 0.77 & 0.65\\
\midrule
\multirow{6}{*}{Forest}
&Retrain & 8.59±1.00 & 10.65±1.18 & 1.25 & 0.27\\
&Stale & 14.64±1.65 & 15.47±1.62 & 1.20 & 0.49\\
&Fine-tune & 9.68±1.14 & 11.45±1.26 & 1.25 & 0.28\\
&NegGrad+ & 10.87±1.10 & 12.71±1.45 & 1.27 & 0.17\\
&NegGrad & 162.41±92.24 & 147.14±38.94 & 0.96 & 0.00\\
&SCRUB & 9.97±1.24 & 11.31±1.35 & 1.25 & 0.29\\

\midrule
\multirow{6}{*}{DMV}
&Retrain & 92.28±41.29 & 60.70±12.69 & 46.93 & 31.32\\
&Stale & 82.33±48.65 & 54.10±7.16 & 40.97 & 42.91\\
&Fine-tune & 104.31±71.19 & 60.22±16.38 & 40.79 & 30.64\\
&NegGrad+ & 66.30±17.51 & 57.98±9.63 & 37.71 & 23.93\\
&NegGrad & 288478.30±152969.35 & 221261.35±119654.25 & 0.01 & 0.00\\
&SCRUB & 157.40±96.48 & 64.61±14.19 & 41.75 & 31.96\\

\midrule

    \end{tabular}}
\end{table}
\begin{table}
\small
  \centering
    \caption{Full and selective deletion for SE/Naru-NeuroCard. 
    Numbers are average relative-error and 95 \% confidence interval for remain queries-QR (full deletion) and queries covering remain and deleted data (selective deletion).}
    \vspace{-0.35cm}
    \label{tab:naru_grouped_ci}
    \resizebox{.95\linewidth}{!}{

    \begin{tabular}{c | c | c c }
\toprule
dataset&model&Full Deletion&Selective Deletion \\
\midrule
\multirow{6}{*}{Census}
&Retrain&16.87±1.08&14.32±1.07\\
&Stale&16.20±0.86&13.81±0.99\\
&Fine-tune&10.23±0.71&9.48±0.79\\
&NegGrad+&10.42±0.70&9.19±0.79\\
&NegGrad&97.62±26.17&107.35±43.48\\
&SCRUB&11.71±0.74&10.93±0.87\\

\midrule

\multirow{6}{*}{Forest}
&Retrain&24.96±4.44&21.66±1.99\\
&Stale&23.78±4.61&20.78±1.71\\
&Fine-tune&21.48±3.67&18.34±1.64\\
&NegGrad+&22.43±4.61&19.38±2.05\\
&NegGrad&74.23±18.83&76.81±17.15\\
&SCRUB&23.12±4.78&19.94±1.86\\

\midrule
\multirow{6}{*}{DMV}
&Retrain&7.37±0.58&8.68±0.61\\
&Stale&13.63±0.61&11.95±1.00\\
&Fine-tune&5.13±0.56&5.77±0.66\\
&NegGrad+&5.34±0.56&9.06±0.70\\
&NegGrad&2892.63±3283.17&40.90±2.63\\
&SCRUB&18.67±1.87&10.25±0.91\\

\midrule
    \end{tabular}}
\end{table}

\begin{table}[]
\small
    \centering
    \caption{Full and selective deletion for DG/TVAE. `retain synth' refers to the f1-score of the xgboost classifier trained with synthetic data and evaluated on the held-out test-set. `delete synth' refers to the f1-score of the xgboost classifier trained with the synth data and evaluate on the deleted rows.}
    \vspace{-0.35cm}
    \label{tab:tvae_grouped}
    \resizebox{.95\linewidth}{!}{
    \begin{tabular}{c | c | c c | c c }
    \toprule
         \multirow{2}{*}{dataset}&\multirow{2}{*}{model}&\multicolumn{2}{c|}{Full Deletion}&\multicolumn{2}{c}{Selective Deletion}  \\
         &&retain synth&delete synth&retain synth&deleted synth \\
         \midrule
         
\multirow{4}{*}{Census}
&Retrain&61.13±0.84&62.84±3.99&61.33±0.93&69.79±0.99\\
&Stale&61.27±0.19&69.33±0.71&60.77±0.95&70.10±0.33\\
&Fine-tune&60.11±1.18&68.84±0.49&61.32±1.18&70.18±0.21\\
&NegGrad+&60.41±0.83&69.09±1.01&61.35±0.65&69.80±1.07\\

\midrule

\multirow{4}{*}{Forest}
&Retrain&41.46±1.00&16.39±0.87&44.57±0.77&21.55±2.66\\
&Stale&46.94±0.80&32.31±1.83&46.15±0.46&32.49±1.68\\
&Fine-tune&48.39±1.49&28.85±2.04&48.89±0.73&34.65±1.07\\
&NegGrad+&49.95±0.78&31.16±1.16&46.32±1.01&33.74±1.13\\
\midrule

\multirow{4}{*}{DMV}
&Retrain&41.69±0.72&36.19±3.76&41.02±0.40&43.64±1.62\\\
&Stale&41.12±0.51&36.99±4.47&41.12±0.48&35.70±5.44\\
&Fine-tune&41.34±0.51&36.08±1.19&40.39±0.29&37.59±5.05\\
&NegGrad+&40.72±0.56&37.40±4.12&41.22±0.72&52.21±6.96\\
\midrule
    \end{tabular}}
\end{table}

\subsubsection{Results using Model-Internal Metrics}
As we have seen above, while investigating the effect of various unlearning methods, looking at only downstream tasks results can be misleading or reveal only part of the full picture. Such results depend on underlying data distributions, on the particular delete operations, and on the actual analytics task at hand. For example, the data subspace affected by a delete and its correlation to the dependent variable may be such that downstream task accuracy results are unaffected, showing all methods (even poor unlearning methods, or even \texttt{Stale}) to perform very well.
In our example tasks studied here, all ML models are generative in nature. Hence, they could generate data items, according to the  distributions they have learned. Then one can compare these distributions against the original (ground truth) data distributions and do so before and after deletions.
In this section, we highlight results using such distributions.
Note that additional model-internal metrics can be utilized. One such example is using a model's likelihood numbers, as we have done above for the AQP/DBEst++ case where the underlying MDN model provides such likelihoods. 
For space reasons, we show these distributions for the AQP/DBEst++ case for the Census dataset in \autoref{fig:dbest_census_full}
(results for the other datasets are very similar). 
In addition to the visual understanding provided by these histograms, we also measure the divergence between the distributions produced by unlearning algorithms and the real distributions after deletions. We use the Jensen-Shannon divergence (a symmetrical version of KL divergence) 
between these distributions. 
JS-divergence values range from 0 to $ln \ 2$ (ca. 0.69). The results are reported in \autoref{tab:js_divergence}. Note that all values are very close to zero, showing strong unlearning performance for the model internally. 

\begin{figure*}[!ht]
\subfloat[]{\label{main:a}\includegraphics[width=.125\textwidth]{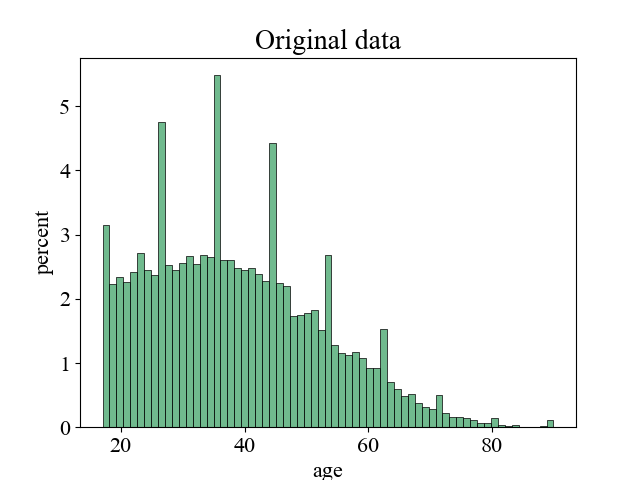}}\hfill
\subfloat[]{\label{main:b}\includegraphics[width=.125\textwidth]{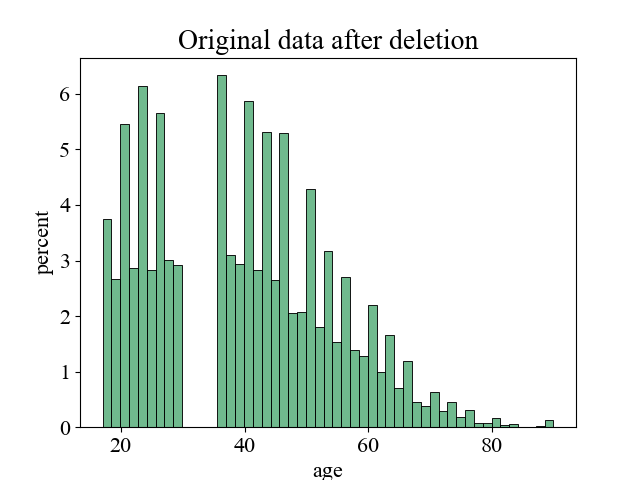}}\hfill
\subfloat[]{\label{main:c}\includegraphics[width=.125\textwidth]{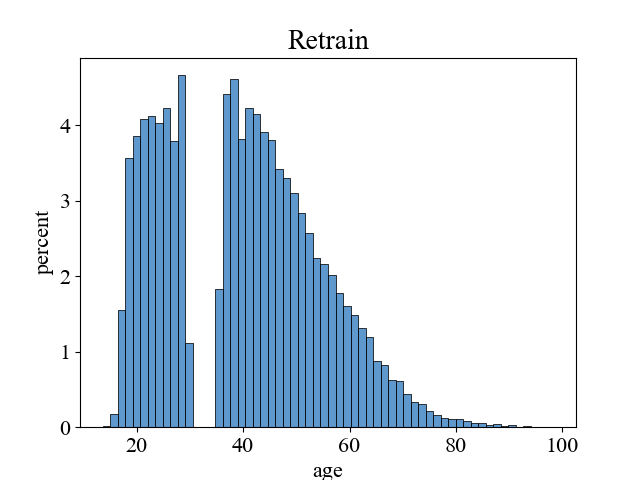}}\hfill
\subfloat[]{\label{main:d}\includegraphics[width=.125\textwidth]{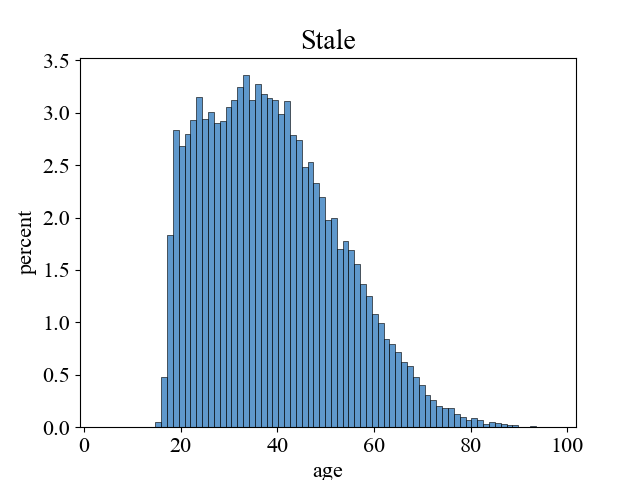}}\hfill
\subfloat[]{\label{main:e}\includegraphics[width=.125\textwidth]{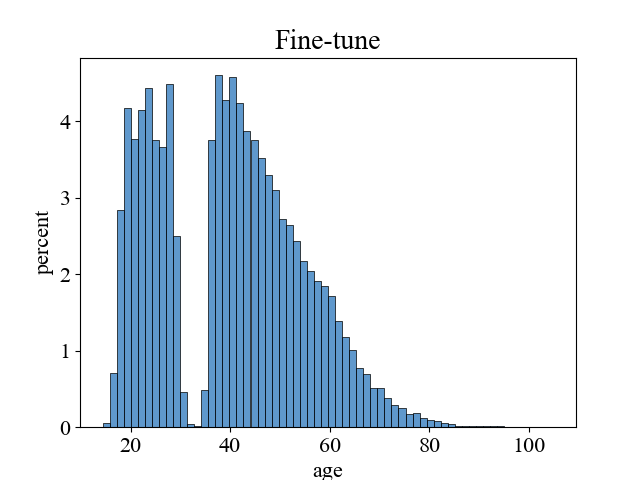}}\hfill
\subfloat[]{\label{main:f}\includegraphics[width=.125\textwidth]{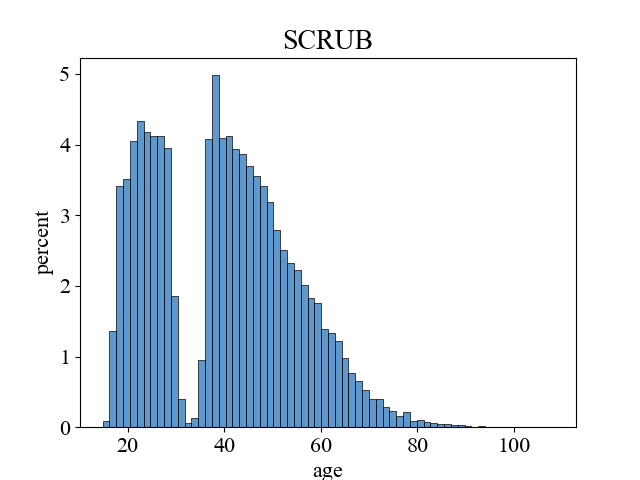}}\hfill
\subfloat[]{\label{main:g}\includegraphics[width=.125\textwidth]{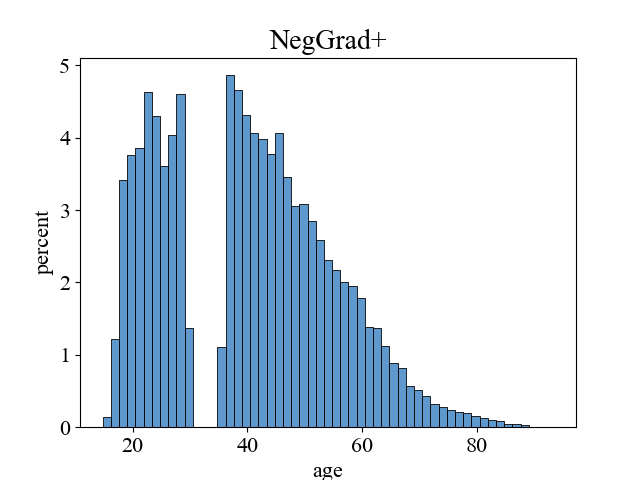}}\hfill
\subfloat[]{\label{main:h}\includegraphics[width=.125\textwidth]{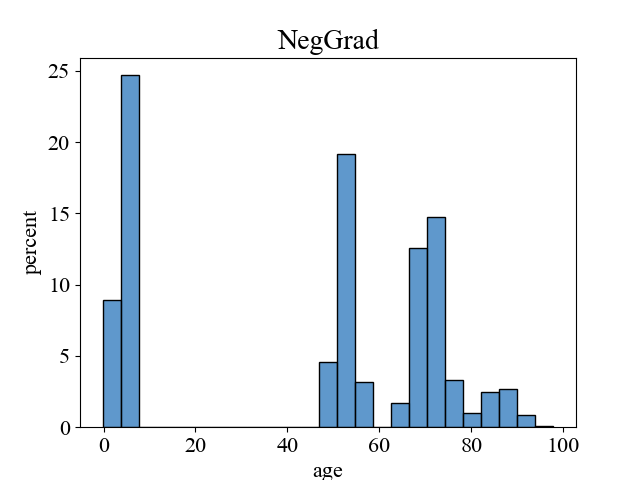}}
\vspace{-0.1cm}
\caption{Histogram of "age" for Census. Native-Country="United States". AQP/DBEst++. "Full" deletion. (a) Original data (b) Original data after deletion (c) \texttt{Retrain} (d) \texttt{Stale} (e) \texttt{Fine-tune} (f) \texttt{SCRUB} (g) \texttt{NegGrad+} (h) \texttt{NegGrad}}
\label{fig:dbest_census_full}
\vspace{-0.1cm}
\end{figure*}

\begin{table}
\small
    \centering
    \caption{ JS divergence between distributions of the original data and the synthetic data generated after deletion. }
    \vspace{-0.35cm}
    \resizebox{.95\linewidth}{!}{%
    \begin{tabular}{c|c|c|c|c|c|c}
    \toprule
         \multirow{2}{*}{method}&\multicolumn{3}{c}{Full}&\multicolumn{3}{c}{Selective} \\
         &Census&Forest&DMV&Census&Forest&DMV \\

    \midrule

Train&0.0156 & 0.0085 & 0.0275 & 0.0096 & 0.0697 & 0.0174 \\
Retrain&0.0118 & 0.0086 & 0.0015 & 0.0053 & 0.0147 & 0.0315 \\
Finetune&0.0198 & 0.0470 & 0.0782 & 0.0019 & 0.0726 & 0.0216 \\
NegGrad+&0.0062 & 0.0255 & 0.0158 & 0.0019 & 0.6407 & 0.0264 \\
NegGrad&0.3084 & 0.3346 & 0.0874 & 0.1001 & 0.6926 & 0.6891\\
SCRUB&0.0278 & 0.0092 & 0.0017 & 0.0015 & 0.0639 & 0.0661 \\

    \midrule
    \end{tabular}}
    \label{tab:js_divergence}
\end{table}

In the results shown in \autoref{fig:dbest_census_full}, the first two subplots show the real data distribution before and after the deletion. 
The other subplots show the distribution of the samples generated by each model after unlearning. The goal is to match the ground truth histogram that is obtained after deletion, shown in subplot (b). One can easily see that \texttt{Stale} does not unlearn, as is expected.
Furthermore, \texttt{NegGrad} is shown to introduce unwanted artifacts in the distribution of the rest of the data as well, showing its limitations. 
All other methods are shown to unlearn the  part of the data that is deleted. 
This scenario and discussion raise the following key issue: As not all deletions will affect downstream task accuracy, one may not wish to apply any unlearning algorithms as downstream accuracy will be unaffected. 
So a ``detector'' for such cases is warranted. Naturally, if the same model is being used for different downstream tasks, such task-specific detectors may not be appropriate, as downstream task accuracy for other tasks may be affected.

\vspace{-0.35cm}
\subsection{Sequential Deletion (Smaller Batches)}
In the above experiments, we performed data deletion in ``one-go''.  
Here, we study a different setting where deletes are executed sequentially (in smaller batches). We aim to address two questions in this section. First, how does accuracy evolve when models are unlearned sequentially? Second, is it better to execute deletion requests on-demand, or to group them and perform unlearning at once?
For these experiments, we use the ``Full'' delete operation described earlier for each dataset. We split it into 5 smaller deletes. At every step, we perform unlearning on the models as updated in the previous step. Figures \ref{fig:dbest_census_sequential} to \ref{fig:naru_census_sequential} show how the median error of downstream tasks evolves in the different settings. 
Overall, three main conclusions can be drawn from these figures. First, \texttt{Fine-tune}, \texttt{NegGrad+} and \texttt{SCRUB} have errors for QR queries that are close to those of \texttt{Retrain}. 
Second, these errors do not accumulate across sequential deletes. \texttt{NegGrad} and \texttt{Stale} conversely show an increase of error in consecutive steps. 
Third, the errors on QD queries  show that \texttt{NegGrad} performs consistently well in terms of forgetting (as expected) along with \texttt{Retrain}, \texttt{Fine-tune}, and \texttt{NegGrad+},  while, \texttt{Stale} again shows a growing error.
These conclusions addressed our first question w.r.t  the evolution of errors. 
 
Figures 
\ref{fig:dbest_retain_ratios} and \ref{fig:naru_retain_ratios} address the second question for AQP/DBEst++ and SE/Naru-NeuroCard for the errors of the queries on the remaining rows. Results for accuracy on deleted rows are very similar and deleted for space reasons.
We also have plotted the horizontal line of 1, to signal the point of no difference between errors in sequential vs the one-go settings. 
The results reveal interesting behaviors. 
First, looking at Figures \ref{fig:dbest_retain_ratios} and \ref{fig:naru_retain_ratios}, we see that for the Census datasets and for AQP/DBEst++ and SE/Naru-NeuroCard there is hardly any difference between sequential vs one-go deletion for all methods. Looking at the Figures for Forest dataset, however, we see emerging differences.
Namely, for AQP/DBEst++ and Forest, one-go deletion for \texttt{Retrain} and \texttt{NegGrad+} emerges as preferable.
And this does not hold for Forest and SE/Naru-NeuroCard (\autoref{fig:naru_retain_ratios}). This highlights the fact that conclusions depend on downstream tasks and datasets.
Interestingly, \texttt{Fine-tune} appears to be less affected than \texttt{Retrain} across these tasks and datasets.
This is important as \texttt{Fine-tune} has proved to be a very strong performer from all previous experiments, competitive to \texttt{Retrain}.
It is obviously also faster than \texttt{Retrain}. And now we see that it appears to be more robust than \texttt{Retrain} with respect to how often it should be run.
So, it appears to be even less expensive.

\begin{figure}
\begin{minipage}{.5\linewidth}
\subfloat[]{\label{main:a}\includegraphics[scale=.25]{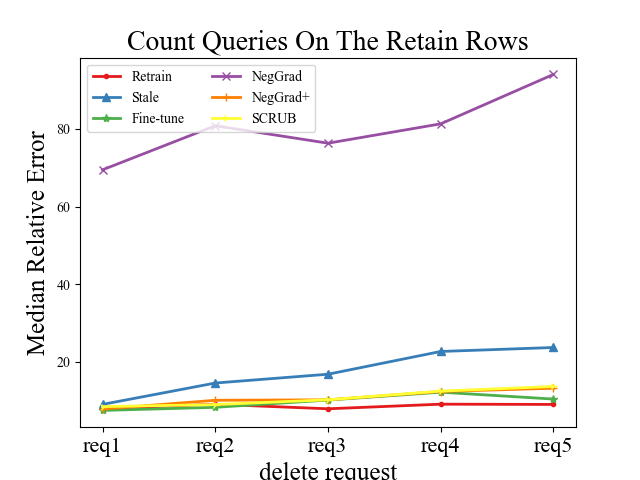}}
\end{minipage}%
\begin{minipage}{.5\linewidth}
\subfloat[]{\label{main:b}\includegraphics[scale=.25]{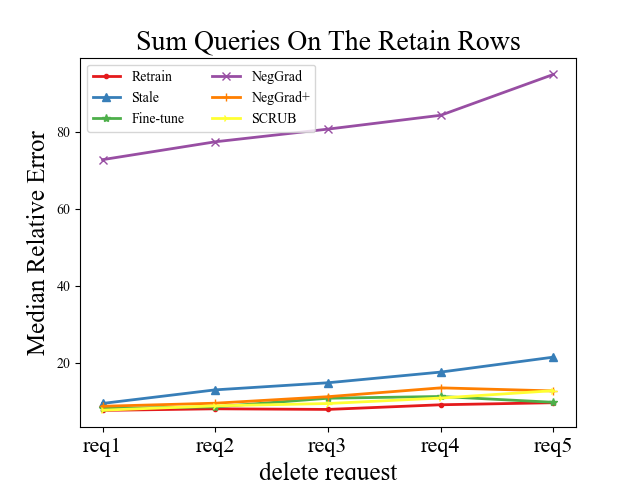}}
\end{minipage} %
\begin{minipage}{.5\linewidth}
\subfloat[]{\label{main:c}\includegraphics[scale=.25]{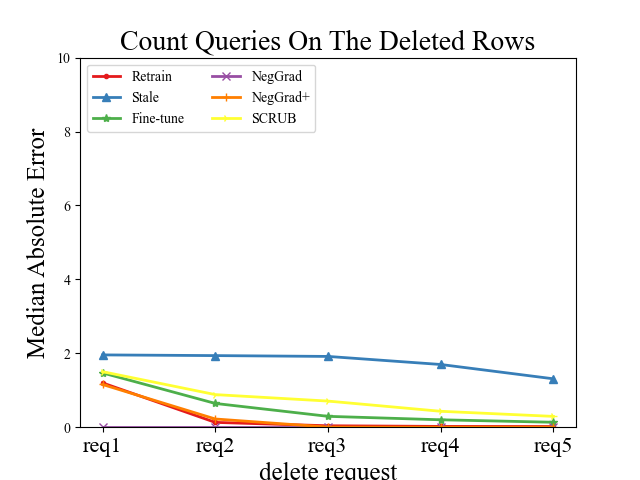}}
\end{minipage}%
\begin{minipage}{.5\linewidth}
\subfloat[]{\label{main:d}\includegraphics[scale=.25]{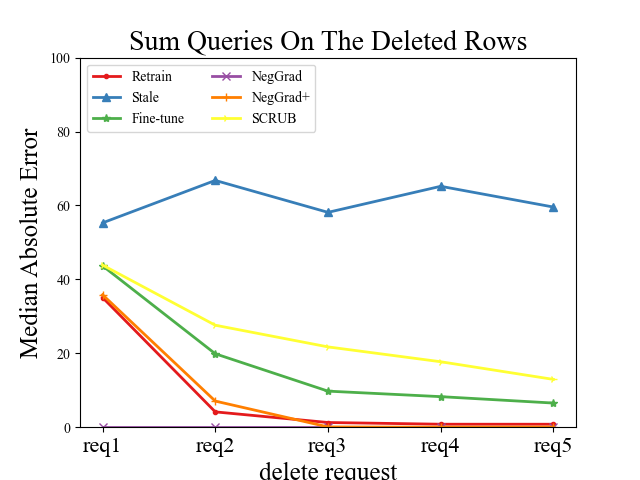}}
\end{minipage} \\%

\vspace{-0.3cm}
\caption{Error evolution. Deleting sequentially. Census dataset, AQP/DBEst++ model. }
\label{fig:dbest_census_sequential}
\vspace{-0.4cm}
\end{figure}
\begin{figure}
\begin{minipage}{.5\linewidth}
\subfloat[]{\label{main:a}\includegraphics[scale=.25]{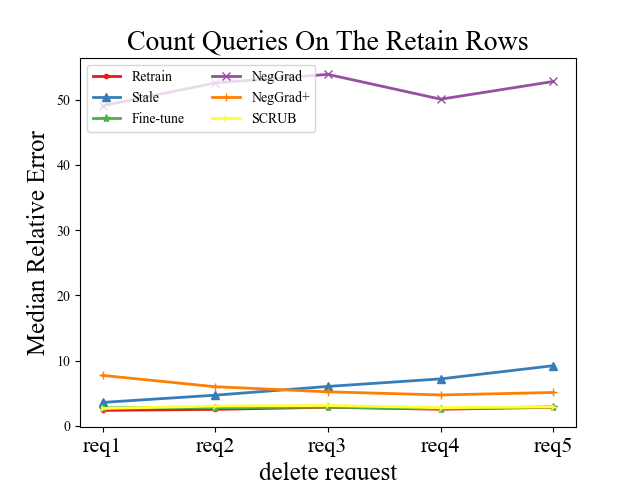}}
\end{minipage}%
\begin{minipage}{.5\linewidth}
\subfloat[]{\label{main:b}\includegraphics[scale=.25]{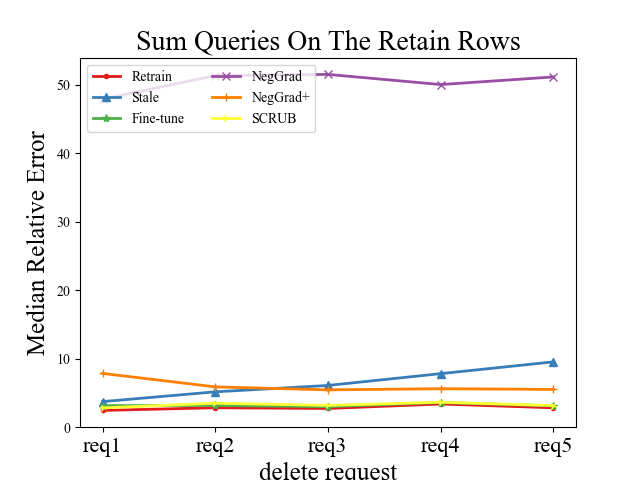}}
\end{minipage} %
\begin{minipage}{.5\linewidth}
\subfloat[]{\label{main:c}\includegraphics[scale=.25]{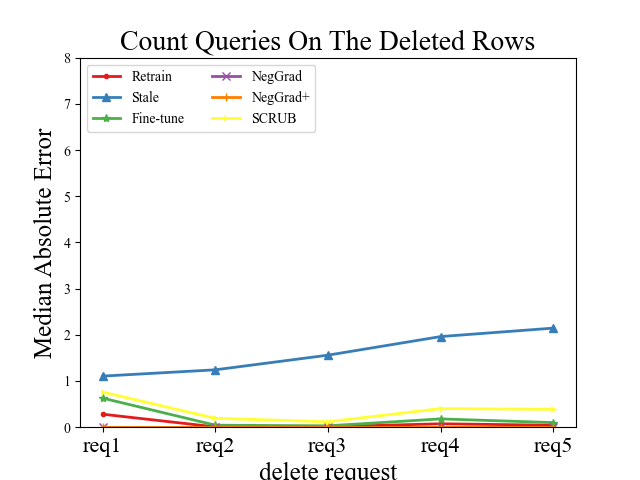}}
\end{minipage}%
\begin{minipage}{.5\linewidth}
\subfloat[]{\label{main:d}\includegraphics[scale=.25]{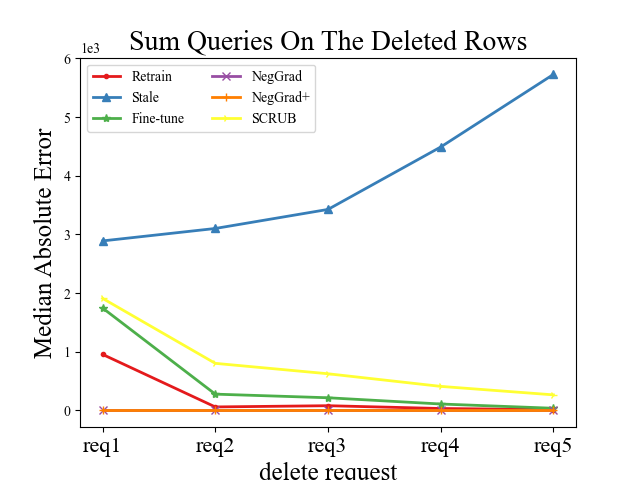}}
\end{minipage} \\%

\vspace{-0.3cm}
\caption{Error evolution. Deleting sequentially. Forest dataset, AQP/DBEst++ model. }
\label{fig:dbest_forest_sequential}
\vspace{-0.4cm}
\end{figure}
\begin{figure}
\begin{minipage}{.5\linewidth}
\subfloat[]{\label{main:a}\includegraphics[scale=.25]{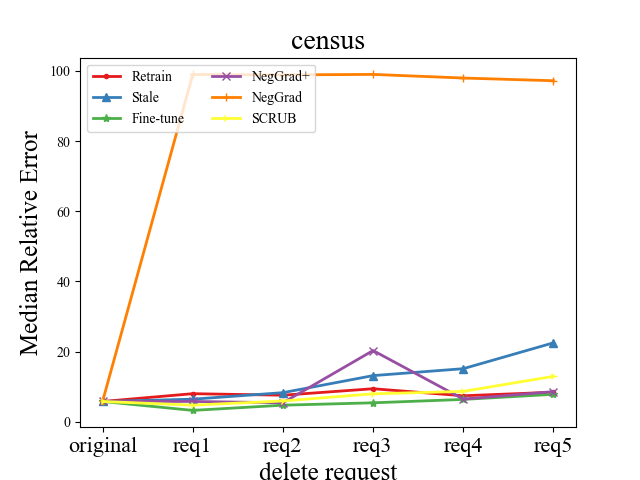}}
\end{minipage}%
\begin{minipage}{.5\linewidth}
\subfloat[]{\label{main:b}\includegraphics[scale=.25]{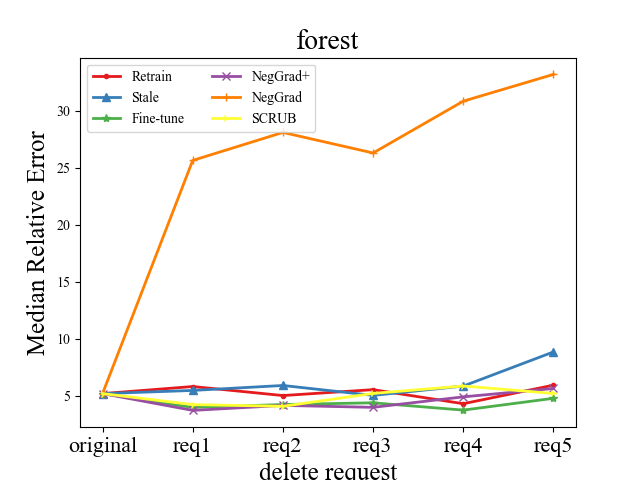}}
\end{minipage} %

\vspace{-0.3cm}
\caption{Error evolution. Sequential deletion. Census dataset, SE/Naru-NeuroCard. }
\label{fig:naru_census_sequential}
\end{figure}

\begin{figure}
\begin{minipage}{.5\linewidth}
\subfloat[]{\label{main:a}\includegraphics[scale=.25]{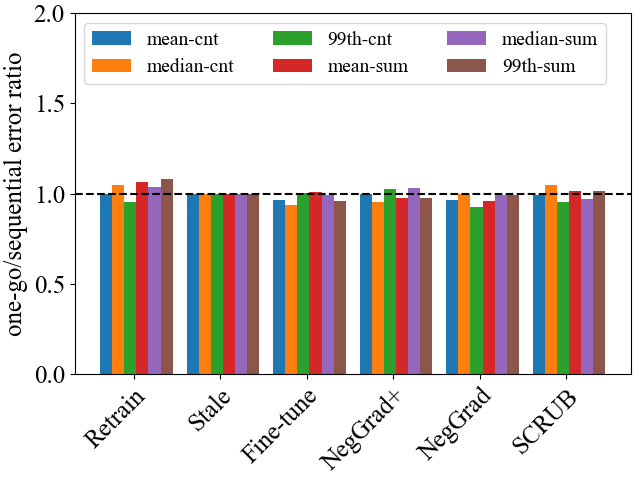}}
\end{minipage}%
\begin{minipage}{.5\linewidth}
\subfloat[]{\label{main:b}\includegraphics[scale=.25]{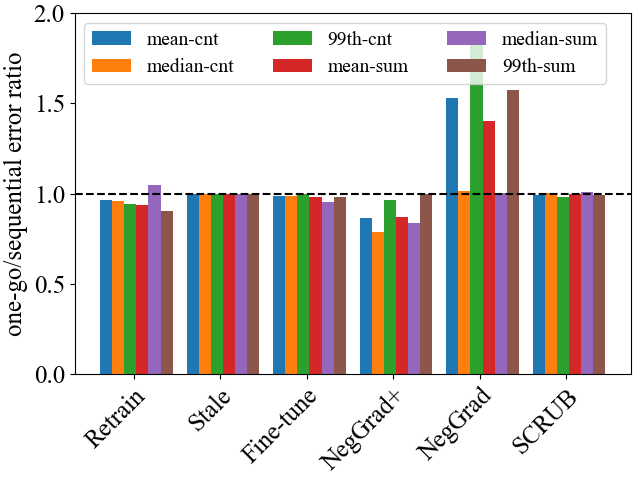}}
\end{minipage} %

\vspace{-0.3cm}
\caption{Ratio of relative errors: deleting in one-go over sequential deletion. AQP/DBEst++ on QR. a) Census, b) Forest. }
\label{fig:dbest_retain_ratios}
\end{figure}

\begin{figure}
\begin{minipage}{.5\linewidth}
\subfloat[]{\label{main:a}\includegraphics[scale=.25]{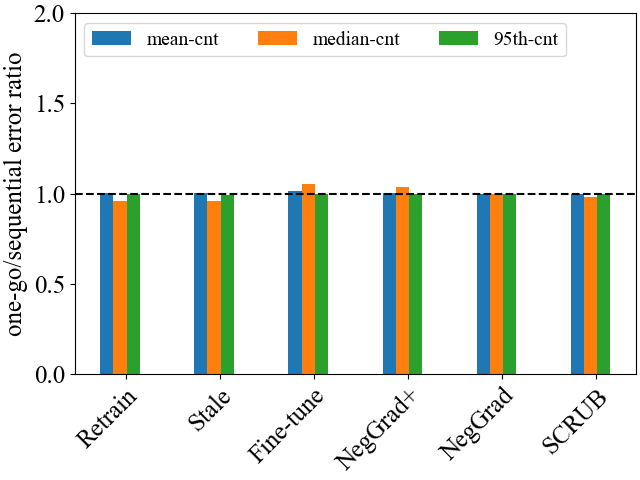}}
\end{minipage}%
\begin{minipage}{.5\linewidth}
\subfloat[]{\label{main:b}\includegraphics[scale=.25]{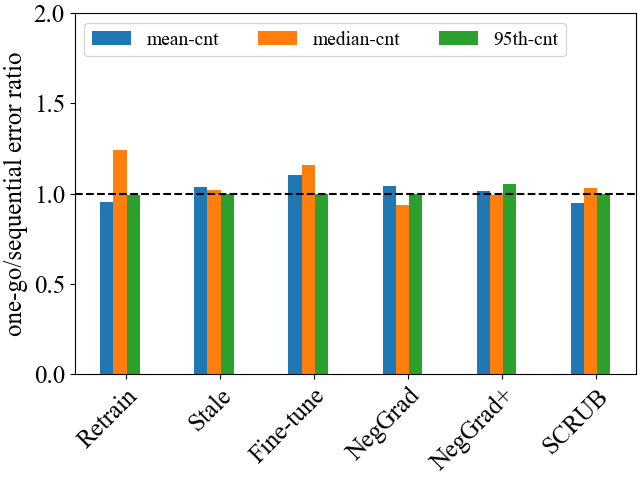}}
\end{minipage} %

\vspace{-0.3cm}
\caption{Ratio of relative errors: deleting in one-go over sequential deletion. SE/Naru-NeuroCard in Forest  on QR. a) Census, b) Forest. }
\label{fig:naru_retain_ratios}
\end{figure}

\subsection{Insertions and Deletions}
Learned DB updatability in general requires support for both data update operations, insertions, and deletions. 
We have shown thus far that \texttt{Fine-tune} (and \texttt{NegGrad+}) perform well for unlearning in a learned DB model. 
However, recent research in learned DB updatability for data insertions in \citep{kurmanji2022detect} showed that fine-tuning is not accurate and more complex algorithms are needed.. 
Thus, it is unknown how these different mechanisms for insertions and deletions would interact with one another.
In other words, can this simple fine-tuning method for unlearning perform well even in conjunction with a method for data insertions? 
We answer this question now  using the code for  \citep{kurmanji2022detect} and combining it with fine-tuning for deletions, as follows: we assume insertions and deletions come in batches. First, we apply the fine-tuning method for deleting a subspace (e.g., all tuples with age values in [20,30]).
Then we apply the method in \citep{kurmanji2022detect} on the fine-tuned model, for inserting back all tuples in the same age group.
Ideally, we would get to a distribution very close to the original, before any deletions and insertions.
\autoref{tab:insert_delete} shows the results. 
After training the original model in Step 1, we perform the deletions in Step 2. Finally, we insert the deleted data again in Step 3.
We show average relative errors for SUM and COUNT queries after each step.
We also report the JS divergence between the original data distribution and generated sample distribution after each step to illustrate that internally, the models at each step stay close to the original data. Note that both errors and JS divergences are very small after Steps 2 and 3.
\begin{table}
\small
  \centering
    \caption{Combining Deletions and Insertions. Numbers show average relative error ± 95 \% CIs. NB: Naru/NeuroCard does not support sum queries (shown as NA). }
   \vspace{-0.35cm}
   \label{tab:insert_delete}
    \resizebox{.95\linewidth}{!}{%
    \begin{tabular}{c | c |c | c c | c}
\toprule
\multirow{2}{*}{Model}&\multirow{2}{*}{Step}&\multirow{2}{*}{Operation}&\multicolumn{2}{c|}{Queries}&\multirow{2}{*}{JS-divergence} \\
&&&Count&Sum& \\
\midrule
\multirow{3}{*}{DBEst++}
&1 & Train & 15.05±1.09 & 15.62±1.08 & 0.0136 \\
&2 & Delete & 17.13±1.39 & 16.00±0.98 & 0.0129 \\
&3 & Insert & 17.10±1.08 & 17.79±1.19 & 0.0044 \\
\midrule
\multirow{3}{*}{NeuroCard}
&1 & Train & 20.85±1.12 & NA & 0.0934\\
&2 & Delete & 10.15±0.71 & NA & 0.1003\\
&3 & Insert&19.50±0.89 & NA & 0.1056\\
\midrule
\end{tabular}}
\end{table}

\subsection{A Data Classification (Upstream) Task}
We report results for the DC task. Following the ML unlearning literature like \citep{kurmanji2023towards}, we perform both class unlearning (delete all examples/tuples of a class) and selective unlearning (deleted only a subset of the class examples/tuples).
For Census, for class unlearning, we remove all tuples of class ' Divorced'. 
And for selective unlearning, we delete only tuples whose age value is between 30 and 40 {\it from all classes}. In both cases, the delete-set amounts to ~10\% of the whole data.
\autoref{tab:tab_classification} shows the classification accuracy for Census - the results for DMV and Forest are very similar.
We can see that Fine-tune continues to enjoy high performance.

\begin{table*}[]
\small
    \centering
    \caption{DC task results with a ResNet architecture. The numbers are average accuracy ± 95\% CIs.}
\vspace{-0.25cm}
    \label{tab:tab_classification}
    \begin{tabular}{c c|c c c | c c c}
    \toprule
         \multirow{3}{*}{Dataset}&\multirow{2}{*}{Method}&\multicolumn{3}{c|}{Class Unlearning} &\multicolumn{3}{c}{Selective Unlearning} \\
         &&Test&Retain&Forget&Test&Retain&Forget \\
\midrule
\multirow{6}{*}{Census}
&Original & 83.69±1.58 &88.65±0.41 &64.72±2.17 
& 83.48±0.60 &85.87±0.39 &81.87±0.84\\
&Retrain & 79.78±1.00 &93.12±0.17 &0.00±0.00 & 84.94±0.51 &86.19±0.44 &80.66±1.12\\
&Fine-tune & 79.03±1.37 &92.27±0.96 &0.00±0.00 & 84.34±1.84 &85.16±1.52 &80.56±4.18\\
&NegGrad & 52.39±8.37 &60.79±1.24 &0.00±0.00 & 1.56±1.51 &1.49±1.81 &0.99±1.61 \\
&NegGrad+ & 76.65±5.83 &89.51±4.13 &0.00±0.00 & 79.51±8.22 &80.80±7.08 &76.66±8.93 \\
&SCRUB & 78.70±0.49 &92.19±0.48 &0.00±0.00 & 83.53±0.47 &85.15±0.41 &80.62±0.87 \\
\midrule
    \end{tabular}
\end{table*}

\subsection{Efficiency}
Unlearning algorithms are motivated by wanting to avoid  the high costs of retrain-from-scratch. 
We thus evaluated the efficiency of the approximate unlearning algorithms studied. 
\autoref{tab:efficiency} shows the speed-up of each algorithm over Retrain for our three downstream tasks and their NN models.
(The results are very similar for the upstream DC task).
We can see that speedups, especially for the \texttt{Fine-tune} and \texttt{Negrad+} are high.
Note that these speedups can become much higher when much larger datasets are used (as training times depend on dataset sizes).

\begin{table*}
\small
    \centering
    \caption{Speed-up of unlearning algorithms over the `Retrain' oracle. Stale is obviously not included.}
\vspace{-0.3cm}
    \resizebox{.95\linewidth}{!}{%
    \begin{tabular}{c|c|c|c|c|c|c|c|c|c|c|c|c|c|c|c|c|c|c}
    \toprule
         \multirow{3}{*}{method}&\multicolumn{6}{c|}{AQP/DBest++}&\multicolumn{6}{c|}{SE/Naru-NeuroCard}&\multicolumn{6}{c}{DG/TVAE} \\
         &\multicolumn{3}{c}{Full}&\multicolumn{3}{c}{Selective}&\multicolumn{3}{c}{Full}&\multicolumn{3}{c}{Selective}&\multicolumn{3}{c}{Full}&\multicolumn{3}{c}{Selective}\\
         &Census&Forest&DMV &Census&Forest&DMV&Census&Forest&DMV &Census&Forest&DMV&Census&Forest&DMV &Census&Forest&DMV \\

    \midrule

    Retrain&1&1&1&1&1&1&1&1&1&1&1&1&1&1&1&1&1&1 \\
    Fine-tune& 11.47&17.05&11.36&28.87&18.37&9.46&1.81&2.00&2.30&2.66&1.94&4.5&16.27&11.86&17.04&19.42&12.57&18.31\\
    NegGrad+& 10.90&10.28&6.67&11.55&10.61&5.70&1.86&2.09&2.28&2.5&2.21&2.45&16.27&15.98&43.08&19.42&15.11&19.88\\
    NegGrad& 10.38&10.23&15.22&12.83&9.69&5.64&2.4&5.86&5.67&5.71&5.22&2.88&NA&NA&NA&NA&NA&NA\\
    SCRUB& 3.63&5.79&1.90&3.72&6.19&1.55&1.56&1.58&1.40&2.15&1.47&2.03&NA&NA&NA&NA&NA&NA\\ 
    \midrule
    \end{tabular}}
    \label{tab:efficiency}
\end{table*}

\subsection{\Rb{SISA Results}} \label{sec:sisa_results}
\Rb{In this section, we summarize \texttt{SISA}'s results, reported in \autoref{tab:sisa_dbest_full} to \autoref{tab:sisa_for_dc}. We only report results for Census dataset due to space, but the conclusions hold for Forest and DMV.}

\Rb{For AQP/DBEst++, \texttt{SISA} consistently demonstrates comparable or improved error rates in comparison to regular training. Retraining \texttt{SISA} generally results in lower errors for deleted queries compared to regular retraining. When examining likelihoods, it's important to note that direct comparisons with regular settings may not be valid due to different model architectures. Nevertheless, the tables reveal that retraining \texttt{SISA} leads to lower average likelihoods for deleted rows, showcasing the impact of unlearning.}

\Rb{\texttt{SISA} shows significantly higher relative error during training compared to regular training in SE/Naru-NeuroCard. Our detailed experiments revealed that constituent models perform better on their own data chunks but struggle with larger data ranges. The introduced sparsity from creating embeddings for the entire table while training on a small shard decreases performance. Additionally, aggregating results across models further increases error. }

\Rb{In DG/TVAE, \texttt{SISA}'s classification performance is comparable to regular training, but retraining during unlearning leads to a slight decrease in accuracy for both retained and deleted sets.}

\Rb{For DC, \texttt{SISA} for training and unlearning has a slight decrease in test and retain accuracy, compared to regular training. This difference persists when comparing regular retraining and SISA retraining. However, surprisingly, the forget accuracy after unlearning SISA is quite similar to the regular retraining from scratch.}

\Rb{Finally, we report \texttt{SISA}'s speed-up both in the training phase compared to the regular training, as well as in the unlearning phase compared to regular retraining. Overall, while training \texttt{SISA} is slower than regular training, during unlearning, retraining \texttt{SISA} is considerably faster than regular retraining. The speed-up is especially remarkably higher in the DC task where we have a higher number of partitions and slices. }

\begin{table*}
\small
  \centering
    \caption{Full/Selective deletion results for AQP/DBEst++ and SISA. Corresponding to \autoref{tab:mdn_full_intervals} and \autoref{tab:mdn_selective_intervals}.}
    \vspace{-0.35cm}
    \label{tab:sisa_dbest_full}
    
    \begin{tabular}{c | c c | c c | c c | c | c c | c c | c}
\toprule
\multirow{2}{*}{model}&\multicolumn{2}{c|}{Query-Retain - QR}&\multicolumn{2}{c|}{Query-Delete - QD}&\multicolumn{2}{c|}{likelihoods}&\multirow{2}{*}{Speedup}&\multicolumn{2}{c|}{Queries}&\multicolumn{2}{c|}{likelihoods}&\multirow{2}{*}{Speedup}\\
&count&sum&count&sum&remain&delete & &sum&count &remain&delete &\\
\midrule

SISA-Train & 14.74±3.15 & 12.92±3.39 &NA&NA&NA&NA    &0.77     & 11.68±2.36 & 11.31±2.66 &NA &NA     &0.77\\
SISA-Retrain & 13.82±3.12 & 15.97±3.91 & 0.00±0.00 & 2.17±1.57 & 8.03 & 0.03   &10.02       & 16.54±2.91 & 13.87±2.73 & 3.88 & 2.96     &9.81\\

\midrule

    \end{tabular}
\end{table*}

\begin{table*}
\small
  \centering
    \caption{The full and selective deletion results for SE/Naru-NeuroCard and SISA. Corresponding to \autoref{tab:naru_grouped_ci}.}
    \vspace{-0.35cm}
    \label{tab:naru_sisa}

    \begin{tabular}{c | c | c c | c }
\toprule
dataset&model&Full Deletion&Selective Deletion & Speedup (full, selective) \\
\midrule
\multirow{2}{*}{Census}
&SISA-Train&62.81±2.95&62.79±2.95    &0.91\\
&SISA-Retrain&48.43±2.31&45.47±2.21     &(8.4, 7.9)\\



\midrule
    \end{tabular}
\end{table*}

\begin{table*}
\small
    \centering
    \caption{Full and selective deletion results for DG/TVAE and SISA. Corresponding to \autoref{tab:tvae_grouped}.}
    \vspace{-0.35cm}
    \label{tab:tvae_sisa}
    \begin{tabular}{c | c | c c | c c | c}
    \toprule
         \multirow{2}{*}{dataset}&\multirow{2}{*}{model}&\multicolumn{2}{c|}{Full Deletion}&\multicolumn{2}{c|}{Selective Deletion}&\multirow{2}{*}{Speedup (full, selective)} \\
         &&retain synth&delete synth&retain synth&deleted synth & \\
         \midrule
         
\multirow{2}{*}{Census}
&SISA-Train&61.81±2.17&NA&NA&NA    &0.84\\
&SISANARetrain&54.09±0.73&18.80±0.11&55.52±1.74&19.81±1.67    &(7.84, 7.01)\\



\midrule
\end{tabular}
\end{table*}

\begin{table*}[]
\small
    \centering
    \caption{SISA for the DC task. Corresponding to \autoref{tab:tab_classification}}
\vspace{-0.25cm}
    \label{tab:sisa_for_dc}
    \begin{tabular}{c |c c c | c c c | c}
    \toprule
         \multirow{2}{*}{Method}&\multicolumn{3}{c|}{Class Unlearning} &\multicolumn{3}{c|}{Selective Unlearning} & \multirow{2}{*}{Speedup} \\
         &Test&Retain&Forget&Test&Retain&Forget & \\
\midrule
SISA-Train & 81.27±0.01 & 87.64±0.01 & 42.31±0.09 & 81.27±0.01 &  81.55±0.01 & 81.07±0.00   &0.81\\
SISA-Retrain & 77.81±0.01 & 90.33±0.00 & 0.00±0.00 & 80.11±0.01 & 80.05±0.02 & 80.72±0.00  &22.01\\
\midrule
    \end{tabular}
\end{table*}

\vspace{-0.25cm}
\subsection{\Rb{Membership Inference Attacks}} \label{sec:mia_results}
\Rb{We perform two types of MIAs to evaluate unlearning quality across baselines. The first is a loss-based attack \citep{kurmanji2023towards}. The second is a confidence-based attack \citep{golatkar2020eternal}. The core idea behind these attacks is to train a binary classifier to learn to distinguish between the members (retain-set or forget-set) and the non-members (validation-set). However, achieving a high accuracy in this classifier doesn't guarantee successful membership detection if the input distributions are very different. To address this, the validation set should follow the distribution of the forget-set, while we must additionally ensure that there is no overlap of rows between those two sets. 
}

\Rb{Applying MIAs to generative models is an underexplored problem since these models do not have a clear output signal like loss or confidence that indicates memorization. Therefore, we only applied the above MIAs to the DC task.}

\Rb{Given the above explanations, we split the table into train, test, and validation sets. Then we randomly unlearn 5\% of the train-set using different unlearning baselines. The results are reported in \autoref{tab:dc_mia}. In both attacks, the attack's accuracy on the original model is higher than the unlearned models, except for \texttt{NegGrad} which results in a very high confidence-based MIA accuracy.}

\begin{table*}[]
\small
    \centering
    \caption{Membership Inference Attack}
    \vspace{-0.35cm}
    \label{tab:dc_mia}
    \begin{tabular}{c | c | c | c | c | c }
    \toprule
    Method & Test Accuracy & Retain Accuracy & Forget Accuracy & loss-MIA & confidence-MIA \\
    \midrule
    Original & 82.11±0.57 &91.15±1.96 & 89.43±2.39 & 57.48±03.18 & 60.64±2.82 \\
    Retrain & 82.48±1.45 & 90.32±2.73 & 72.18±3.08 & 49.25±02.91 & 54.45±3.72 \\
    Fine-tune & 83.84±0.91 & 85.68±0.32 & 76.40±1.00& 51.31±04.10 & 52.22±3.32 \\
    NegGrad & 8.32±4.76 & 8.68±5.38 & 7.39±3.75 & 52.52±02.82 & 66.60±78.10 \\
    NegGrad+ & 72.72±26.68 & 72.82±25.75 & 63.52±24.86 & 50.00±03.36 & 48.59±11.74 \\
    SCRUB & 84.29±0.65 & 84.44±0.27 & 76.27±3.09 & 47.10±01.99 & 49.12±5.03 \\
    SISA-Train & 83.32±0.27 & 82.92±0.12 & 76.05±0.64 & 49.60±0.21 & 57.40±0.34 \\
    SISA-Retrain & 83.04±0.22 & 82.72±0.37 & 75.40±0.87 & 49.60±0.66 & 57.72±0.46 \\
    \midrule
    \end{tabular}
\end{table*}

\vspace{-0.35cm}
\subsection{\Rb{Trade-offs}} \label{sec:trade_off}
\Rb{Machine Unlearning algorithms trade off three aspects: model utility, forget quality, and speed. In our experiments, model utility is quantified as the models' performance (accuracy) on downstream tasks. Forget quality is captured using likelihood, sample distribution, MIA, as well as accuracy on the forget set. 
Our results throughout illustrate what axes of these trade-offs are improved or sacrificed by each unlearning algorithm. }

\Rb{In the exact unlearning algorithms, while \texttt{Retrain} does a perfect job of maintaining utility and forgetting at the same time, it could be very slow, as evidenced by the speedups associated with some approximate unlearning algorithms. The \texttt{SISA} version for exact unlearning can also unlearn quickly and with high quality, but the utility is not always guaranteed (as is the case for SE/Naru-NeuroCard (e.g., \autoref{tab:naru_sisa}))}. 

\Rb{For approximate unlearning algorithms, while all of them show speed-ups over \texttt{Retrain}, their utility vs. forgetting trade-offs are different. \texttt{NegGrad} usually does a top job of unlearning w.r.t. different metrics  (likelihoods and accuracy on the deleted data). However, it catastrophically damages utility (e.g.,  \autoref{tab:mdn_full_intervals}).  \texttt{NegGrad+} and \texttt{SCRUB} show better trade-offs between forgetting, utility and speed-up over \texttt{Retrain}. \texttt{Fine-tune}, on the other hand, consistently performs well by efficiently forgetting, without damaging utility, and while offering significant speedups over \texttt{Retrain}.}

\Rb{Comparing \texttt{SISA} exact unlearning vs the top performers of approximate unlearning, we note the following: First, in general, \texttt{SISA} speedups highly depend on how many of its models need to be retrained. This, in turn, depends on how data partitions are defined and how the forget set at any given time is distributed over these partitions. Our results show several cases where \texttt{Fine-tune} speedups are significantly higher than \texttt{SISA}'s. 
With respect to model utility also, for some tasks, \texttt{SISA} may underperform, likely owing to the aggregation of involved models whereby errors may accumulate. For forgetting quality, \texttt{SISA} is generally a top performer.}

\section{Lessons Learned}

{\it Is machine unlearning in learned DBs different than updating learned DBs with new data insertions?}
Our investigation has revealed an interesting, perhaps surprising finding: when it comes to deleting data in learned database systems, with commonly-used models on commonly-studied tasks, simple methods (like \texttt{Fine-tune} and \texttt{NegGrad+}) do very well.
This finding is interesting because it is in stark contrast with two other observations: First, in the context of learned database systems, \textit{inserting} data is a hard problem and simple fine-tuning approaches evidently fail badly \citep{kurmanji2022detect}. 
When it comes to the first discrepancy, one hypothesis is that, in the current scenarios studied in learned databases, inserting data is a harder problem than deleting data. One contributing factor to this is that the new ``knowledge'' that is inserted may interfere with old ``knowledge''. Simple methods, like \texttt{Fine-tune}, interfere with old knowledge by fine-tuning for new data, while ''catastrophically forgetting'' old data.
The issue of catastrophic forgetting in the machine learning community is a challenging one and requires dedicated solutions that are more sophisticated than simple fine-tuning. Instead, removing knowledge does not face this difficulty. However, it is still an open problem to identify whether there are other unique difficulties associated with deleting knowledge from learned components of database systems that were not surfaced in our initial investigation into the topic. For instance, a systematic understanding of how different aspects of the deletion problem (e.g. size and homogeneity of delete-set, relationship between delete-set and retain-set) affect results is an open problem for future research.

{\it Is machine unlearning in learned DBs different than machine unlearning in image classification?}
In the context of image classification (IC), which is the common testbed for unlearning algorithms in the ML community, the simple approach of fine-tuning yields poor results \citep{golatkar2020eternal,golatkar2020forgetting,kurmanji2023towards}. 
And, interestingly, (our adaptation of) \texttt{SCRUB}, a top-performing unlearning method for IC substantially outperforming \texttt{Fine-tune} (and \texttt{NegGrad+}) on IC tasks, does not generally do better than \texttt{Fine-tune} for our DB tasks.


An important difference between the setup we study in this paper compared to the standard unlearning benchmarks in IC tasks is that the latter uses significantly deeper neural networks, with significantly more trainable parameters. Given this, we hypothesize that perhaps the success of simple approaches like fine-tuning in our case is (at least partially) due to the shallower nature of the neural networks we use. More concretely, we have 2, 5, and 4 hidden layers in our networks for the MDN in AQP/DBEst++, for the DARN in SE/Naru-NeuroCard, and for the TVAE in DG, respectively. While, for instance, \citep{kurmanji2023towards} uses models with more than 20 hidden layers for image classification. To investigate this, we ran an experiment on the computer vision tasks with shallower models than those typically used (in Table \ref{tab:vision_results}), and indeed we observed that this intervention led to a significantly reduced gap in the performance of the state-of-the-art algorithms over fine-tuning. This is an important finding for this community, as it indicates that transitioning to using deeper networks in the future may come with tough growing pains for data deletion in learned data systems. It is also an important finding for ML unlearning research, as it ties unlearning performance to characteristics of the network architecture (shallower vs deeper) - a connection not previously made.

\begin{table*}
\small
    \centering
    \caption{Results for image classification (IC) on Cifar5 and Lacuna5 
    datasets. There are 5 classes and 100 samples in each class \citep{golatkar2020eternal}. We unlearn class 0. An All-CNN network \citep{springenberg2014striving} with 3 CNN layers (much shallower than the usual models for IC tasks, which have 20+ layers). 
    Fine-tune and SCRUB perform similarly in terms of delete-error -- a very different conclusion compared to the IC results with deeper networks \citep{kurmanji2023towards}. In terms of test- and retain-error, the results are mixed. This is an important finding: For shallower networks, the effect of more sophisticated unlearning algorithms is much less pronounced. 
    For shallower networks, for IC tasks \texttt{Fine-tune} does as well as the others across 3 errors. But, for these shallow networks test and retain-errors are very bad, so deeper networks are needed.}
    \label{tab:vision_results}
\vspace{-0.35cm}
    \begin{tabular}{c | c c c | c c c }
    \toprule
         \multirow{2}{*}{model}&\multicolumn{3}{c|}{Cifar5}&\multicolumn{3}{c}{Lacuna5} \\
         &test-error&retain-error&delete-error&test-error&retain-error&delete-error\\
         \toprule
         original&42.75±1.1&37.75±0.7&71.0±1.9&49.2±1.2&43.5±0.9&54.0±1.0\\
         \midrule 
         Retrain&43.0±1.3&30.2±0.0&100.0±0.0&46.5±1.4&44.7±0.0&100.0±0.0\\
         Fine-tune&40.5±0.7&30.5±0.8&98.0±1.0&41.7±1.8&39.5±2.1&100.0±0.0\\
         SCRUB&44.5±2.1&39.7±1.8&97.0±0.8&37.0±1.4&35.7±1.5&100.0±0.0\\
         \midrule
    \end{tabular}
\end{table*}

{\it On internal-model accuracy and downstream task accuracy for unlearning.} 
We have also learned that looking at only downstream task accuracy may be misleading. As our experiments have shown, downstream accuracy is very much dependent on dataset characteristics, delete operation characteristics, relationships between deleted data and retained data, and the downstream task itself. In some cases, even \texttt{Stale} performs well, for example. It is instructive, therefore, to use model-internal metrics (like likelihood, in cases the ML models provide them, and generated learned distributions of ML models and related distribution-distances like JS divergence) in addition which can help explain downstream task accuracy.
Looking at such internal-model metrics is also very valuable in any experimental analysis, where inevitably only a select set of experiments are (can be) performed. Thus, looking only at downstream task accuracy may be dependent on data and/or model and/or delete operation peculiarities and may be misleading.
 
{\it On accuracy, overheads, and frequency of unlearning.}
The ideal unlearning method would ensure high accuracy on retained and deleted data at very low overheads.
This is the raison d' etre of this research field; that is, to avoid running \texttt{Retrain} which (especially for very large datasets) can be prohibitively expensive. The experiments of sequential vs one-go deletions showed that different methods have different sensitivities on datasets and downstream tasks. \texttt{Fine-tune} appears to be less sensitive to these, affording the luxury of running it less frequently (i.e. on larger batches of deletion requests). 
So \texttt{Fine-tune} (and \texttt{NegGrad+}), in addition to their accuracy being very competitive under all tasks/models and datasets studied, and to the fact that they are very efficient methods (high speedups versus Retrain), they can also be run less frequently, as errors are shown not to be accumulating over time.

{\it On combining algorithms for continuous learning and unlearning.}
The simplicity of \texttt{Fine-tune} as an unlearning algorithm does not adversely interact with more complex algorithms needed to ensure continuous learning, as new data insertions occur. We have seen that these can be combined (even deleting and inserting data in overlapping data spaces) providing a comprehensive highly accurate solution for both continuous learning and unlearning.

\section{Conclusion}
We have presented the first study of unlearning in learned database systems. 
This is a crucial ingredient for successfully updating models in NN-based learned DBs in the face of frequent data updates that characterize  DBs. 
And is the only ingredient currently completely lacking thorough research and findings.
Our  investigation covered three different downstream tasks (AQP, SE, and DG), each employing a different generative neural-network-based model, and one upstream task based on a discriminative NN model (DC), and across three different datasets. 
It studied different unlearning methods, ranging from simple baselines, such as \texttt{Fine-tune} and \texttt{NegGrad},  more sophisticated methods, such as \texttt{NegGrad+}, and a state-of-the-art unlearning algorithm, \texttt{SCRUB}, adapted from the machine unlearning literature.
Our investigation proposed and studied appropriate metrics including downstream-task specific, as well as model-internal specific metrics in order to substantiate and interpret results.
Our results answer a large number of related key questions with respect to key learned DBs. 
They also point to different conclusions compared to those from research in insertion updates in learned DBs and from the ML community's findings for unlearning in image classification tasks.
The work puts forth the basic skeleton (for instance, unlearning algorithm baselines, performance metrics, and downstream and  upstream tasks) as well as related findings en route to a much-needed benchmark for machine unlearning in learned DBs.



\bibliographystyle{ACM-Reference-Format}
\bibliography{main}

\end{document}